\documentclass[lettersize,journal]{IEEEtran}

\usepackage{array}
\usepackage{textcomp}
\usepackage{stfloats}
\usepackage{verbatim}
\usepackage{cite}

\usepackage{graphicx}

\usepackage{float}
\usepackage{placeins}
\usepackage{booktabs}
\usepackage{outlines}
\usepackage{epsfig}
\usepackage{amsmath,amssymb,amsfonts}
\usepackage{subfigure}
\usepackage{xcolor}
\usepackage{threeparttable}
\PassOptionsToPackage{hyphens}{url}
\usepackage[hyphens]{url}
\usepackage[bottom]{footmisc}
\usepackage{multirow}
\usepackage{xspace} 
\usepackage{lipsum}
\usepackage{amsfonts} 
\DeclareGraphicsExtensions{.png, .pdf, .jpg, .bmp}
\usepackage{tikz}
\usepackage{tabularx}
\usepackage{multirow}
\usepackage{balance}
\usepackage{makecell}
\usepackage{adjustbox}
\usepackage{soul}
\usepackage{pifont}

\usepackage{algorithm}
\usepackage[noend]{algpseudocode}

\usepackage{enumitem}
\setlist{topsep=1pt,itemsep=1pt,parsep=1pt,itemindent=0pt,leftmargin=0.13in}
\usepackage{outlines}
\usepackage{soul}

\newcommand{\eBPF}{\textit{E-MiddleNet}\xspace}
\newcommand{\dpdk}{\textit{D-MiddleNet}\xspace}
\newcommand{\skmsg}{\texttt{SKMSG}\xspace}
\newcommand{\xsk}{\texttt{XSK}\xspace}

\newcommand{\name}{MiddleNet\xspace}

\newcommand{\ie}{{\em i.e., \/}}
\newcommand{\eg}{{\em e.g., \/}}
\newcommand{\etc}{{\em etc. \/}}

\newcommand{\hide}[1] {}

\definecolor{majorelleblue}{rgb}{0.38, 0.31, 0.86}

\newcommand{\revdelete}[1]{}

\begin{document}

\title{\name: A Unified, High-Performance NFV and Middlebox Framework with eBPF and DPDK}

\author{
    Shixiong Qi,
    Ziteng Zeng, 
    Leslie Monis, and
    K. K. Ramakrishnan,~\IEEEmembership{Fellow,~IEEE,} \\
    Dept. of Computer Science and Engineering, University of California, Riverside
}

\markboth{UCR CSE Networking Group Technical Report, Net-2023-0307}%
{Shell \MakeLowercase{\textit{et al.}}: A Sample Article Using IEEEtran.cls for IEEE Journals}


\maketitle

\begin{abstract}
Traditional network resident functions (e.g., firewalls, network address translation) and middleboxes (caches, load balancers) have moved from purpose-built appliances to software-based components. However, L2/L3 network functions (NFs) are being implemented on Network Function Virtualization (NFV) platforms that extensively exploit kernel-bypass technology. They often use DPDK for zero-copy delivery and high performance. On the other hand, L4/L7 middleboxes, which have a greater emphasis on functionality, take advantage of a full-fledged kernel-based system. 

L2/L3 NFs and L4/L7 middleboxes continue to be handled by distinct platforms on different nodes. 
This paper proposes \name that develops a unified network resident function framework that supports L2/L3 NFs and L4/L7 middleboxes.
\name supports function chains that are essential in both NFV and middlebox environments.
\name uses the Data Plane Development Kit (DPDK) library for zero-copy packet delivery without interrupt-based processing, to enable the `bump-in-the-wire' L2/L3 processing performance required of NFV. 
To support L4/L7 middlebox functionality, \name utilizes a consolidated, kernel-based protocol stack for processing, avoiding a dedicated protocol stack for each function. \name fully exploits the event-driven capabilities of the extended Berkeley Packet Filter (eBPF) and seamlessly integrates it with shared memory for high-performance communication in  L4/L7  middlebox function chains. The overheads for \name in L4/L7 are strictly load-proportional, without needing the dedicated CPU cores of DPDK-based approaches.
\name supports flow-dependent packet processing by leveraging Single Root I/O Virtualization (SR-IOV) to dynamically select the packet processing needed (Layers 2 - 7). 
Our experimental results show that \name  achieves high performance in such a unified environment.\footnote{\textit{
This paper is an extended version of our previously published IEEE NetSoft 2022~\cite{middlenet-netsoft2022} paper and IEEE TNSM paper~\cite{middlenet-tnsm}.
In this extended version, we additionally perform overhead auditing of our shared memory-based design (\S\ref{sec:shm-design}) to clearly show the reason why shared memory communication can fundamentally improve the data plane performance for a chain of L2/L3 NFs or L4/L7 middleboxes (details in Appendix-\ref{sec:shm-auditing}). 
}}
\end{abstract}

\begin{IEEEkeywords}
Middleboxes, NFV, DPDK, eBPF, service function chains.
\end{IEEEkeywords}

\vspace{-1mm}\section{Introduction}\vspace{-1mm}

Networks have increasingly become software-based, using virtualization to exploit common off-the-shelf (COTS) hardware to provide a wide array of network-resident functions, thus avoiding having to deploy functions in purpose-built hardware appliances.
This has broadened the networking capabilities provided by both the network and cloud platforms, offloading the burden from end-hosts that may have limited power and compute capability (\eg cell phones or IoT devices). With software-based network-resident functions, network services can be more agile. They can be deployed more dynamically on end-systems that house multiple services.

But there continues to be a dichotomy in how various network-resident services are supported on software-based platforms. Layer 2 and Layer 3 (L2/L3) functions that seek to be transparent and act as a bump-in-the-wire are currently being supported with Network Function Virtualization (NFV) technologies. These focus on performance and are built with network functions (NFs) running in userspace supported by {\it kernel-bypass} technology such as Data Plane Development Kit (DPDK~\cite{dpdk}). Primarily providing switching (demultiplexing and forwarding), they typically do not provide a full network protocol stack, and are exemplified by approaches such as OpenNetVM~\cite{opennetvm} and OpenvSwitch (OVS)~\cite{ovs}.

On the other hand, middleboxes operating at Layer 4 through Layer 7 (L4/L7) require the {\it full network protocol stack's processing} (\eg for application layer functionality such as HTTP proxies), in addition to more complex stateful functionality in userspace, including storage and other I/O operations (\eg caching). Thus, flexibility and functionality are prominent concerns, with performance being a second (albeit important) consideration. A robust and proven kernel-based protocol stack is often desirable~\cite{Kenichi2016StackMap}, as specialized userspace protocol stack implementations often do not support all possible corner cases. 

These distinct requirements for NFV and middlebox designs typically result in the need for different systems. However, networks require both types of functionality to be supported concurrently for different flows, and in many cases, even for the same flow. This calls for supporting them in a {\it unified} framework so that they can be deployed on COTS end-systems dynamically and flexibly. 

Both NFV and middleboxes often have to build complex packet processing pipelines using function chaining. This helps ease development through the use of microservices, which can be independently scaled as needed to improve resource utilization. But the excessive overhead (\eg interrupts, data copies, context switches, protocol processing, serialization/deserialization) incurred within the data plane of current service function chains can be a deterrent.
Even worse, the data plane overhead in current function chaining solutions increases with the function chain size, which significantly reduces their data transfer performance (see \S\ref{sec:auditing}).

Using shared memory communication can help us achieve a more streamlined, efficient data plane design. Shared memory communication supports zero-copy packet delivery between network-resident functions, by having a shareable backend buffer to store packet data, avoiding unnecessary data plane overheads within a function chain. 

Another dichotomy is in how the key building block for shared memory communication is designed. This relates to {\it how packets are moved between the NIC and the shared memory buffer}, and {\it how packet descriptors are passed between functions in a function chain}.
The first option is to exploit the \textit{\textbf{event-driven}}
networking subsystem provided by the extended Berkeley Packet Filter (eBPF~\cite{ebpf}). eBPF offers extensive toolkits (\eg AF\_XDP~\cite{afxdp}, \skmsg~\cite{skmsg}) in support of zero-copy packet delivery.
Importantly, eBPF incurs negligible overhead in the absence of events (such as packet arrivals to a given function or even to the platform), making it an excellent fit for supporting a rich set of diverse,  efficient network-resident functions.
An eBPF program does have size restrictions and must run to completion, requiring careful design~\cite{miano2018creating}. A second alternative approach is to build the shared memory communication framework around \textit{\textbf{polling-based}} DPDK, as has been used in many high-performance virtualized software-based networking environments, \eg OpenNetVM~\cite{opennetvm}. They provide zero-copy delivery into the userspace. Using poll-mode drivers (PMD)~\cite{dpdk-pmd} and RTE RING~\cite{dpdk-ring_lib}, they avoid the deleterious effects of interrupt-based processing of network I/O (\eg receive-livelocks) under overload~\cite{mogul1997eliminating}, making it possible to support complex function chaining at line rate. Nevertheless, dedicated polling continuously consumes  significant CPU resources, and thus is not load-proportional. While this may be reasonable in an NFV-only dedicated system, it is challenging for systems that host many services, including middlebox functions.

In this work, we develop \name, a unified, high-performance NFV and middlebox framework. We take a somewhat unconventional approach by examining an event-driven eBPF design, and separately a polling-based DPDK design for supporting NFV and middlebox function chains with shared memory, and evaluating each design approach.
We then arrive at the design of \name as the most suitable framework for a unified platform supporting both NFV and middlebox functionality. \name uses Single Root I/O Virtualization (SR-IOV~\cite{Dong2010SRIOV}) to enable their co-existence.

\name makes the following contributions:

\noindent\textbf{(1)} We qualitatively discuss the usability of different data plane models for supporting NFV and middlebox capabilities. We carefully audit their data plane overheads and quantitatively assess the performance of each approach. We also look at how current data plane models support function chaining (\S\ref{sec:background}).

\noindent\textbf{(2)} We then design the shared memory communication for \name both the NFV and middlebox (\S\ref{sec:shm-design}) functionality. We (qualitatively and quantitatively) examine the suitability of eBPF and DPDK in supporting different aspects of shared memory communication, including NIC-shared memory packet exchange and zero-copy I/O (\ie packet descriptor delivery) within the function chain (\S\ref{sec:l2l3-middlenet} and \S\ref{sec:l4l7-middlenet}). This helps us understand the strengths and limitations of each option (DPDK's PMD, polling/interrupt-based AF\_XDP in eBPF, DPDK's RTE RING, eBPF's \skmsg), and the root causes. \name chooses to leverage the strengths of polling-based DPDK for L2/L3 NFV, and takes advantage of event-driven eBPF for L4/L7 middleboxes, to strike the balance between performance and resource efficiency.

\noindent\textbf{(3)} For achieving a unified NFV/middlebox framework, we evaluate different alternatives: a hardware-based approach (via SR-IOV~\cite{Dong2010SRIOV}) and a software-based approach  (via virtual device interfaces, \eg virtio/vhost~\cite{virtio-vhost-net}). We assess the performance with SR-IOV and recommend its use for the unified design because of its minimum data plane overhead (\S\ref{sec:unified}).

\noindent\textbf{(4)}  
\name supports function-chain-level isolation to address security concerns with shared memory communication. We create a private memory pool for each function chain to prevent unauthorized access from untrusted functions outside the chain. \name further enhances traffic isolation by applying packet descriptor filtering between functions (\S\ref{sec:security-domain}).

\vspace{-1mm}\section{Background and Motivation}\label{sec:background}\vspace{-1mm}

We examine a number of virtualization frameworks and the networking support that can be provided for supporting network resident functions. We audit the data plane overheads for these different combinations of virtualization frameworks and networking approaches, and discuss their applicability for achieving a high-performance, lightweight, and unified NFV/middlebox framework.

\vspace{-1mm}\subsection{Basic elements in supporting network resident functions}\label{sec:basic}\vspace{-1mm}

\begin{figure*}[htbp]
\centering
    \includegraphics[width=\textwidth]{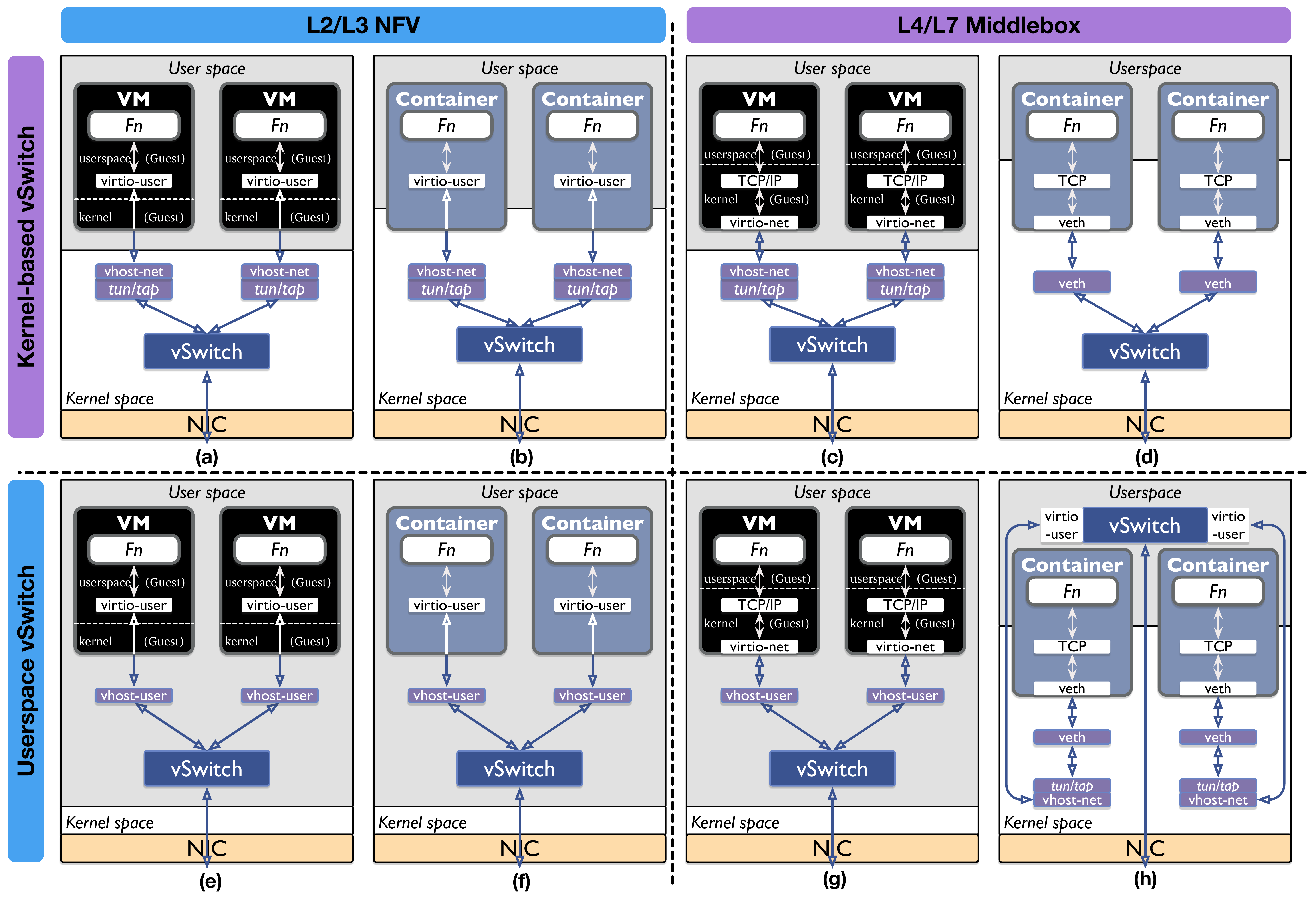}\vspace{-5mm}
\caption{Distinct data plane models for NFV and Middlebox, with different vSwitch options, virtual device interfaces, and virtualization frameworks: (a) kernel-based vSwitch + \textit{virtio-user/vhost-net} \& \textit{TUN/TAP} + VM;  (b) kernel-based vSwitch + \textit{virtio-user/vhost-net} \& \textit{TUN/TAP} + container; (c) kernel-based vSwitch + \textit{virtio-net/vhost-net} \& \textit{TUN/TAP} + VM; (d) kernel-based vSwitch + \textit{veth} + container; (e) userspace vSwitch + \textit{virtio-user/vhost-user} + VM; (f) userspace vSwitch + \textit{virtio-user/vhost-user} + Container; (g) userspace vSwitch + \textit{virtio-net/vhost-user} + VM; (h) userspace vSwitch + \textit{virtio-user/vhost-net} \& \textit{TUN/TAP} + \textit{veth} + container.
\textbf{We assess (f) as the best solution for L2/L3 NFs and (d) as the best solution for L4/L7 middleboxes (\S\ref{sec:auditing}).}}
\label{fig:ovs-datapath}\vspace{-5mm}
\end{figure*}

We identify four key elements for building NFV and middleware environments, including \textbf{virtualization frameworks}, the \textbf{virtual switch (vSwitch)}, the \textbf{protocol stack}, and the \textbf{virtual device interface}. Virtualization helps to multiplex compute resources, and can greatly improve resource efficiency, and reduce costs, while also providing isolation for building L2/L3 NFs and L4/L7 middleboxes. A vSwitch is typically used to provide L2 forwarding/L3 routing. The network protocol stack, often implemented in the OS kernel, provides protocol layer processing (\eg TCP/IP). It is necessary for L4/L7 middleboxes, but is less important for L2/L3 NFs. Virtual device interfaces are used to connect the virtualized function and its protocol stack (for L4/L7 middleboxes only) to the vSwitch, thus building a complete NF and middlebox environment. There are several alternatives for each of these elements, which we describe below.

\noindent\textbf{Virtualization frameworks:} Widely-adopted virtualization frameworks include virtual machines (VMs) and containers. 
VMs often depends on hardware-level virtualization supported by the Virtual Machine Monitor (VMM) or the hypervisor in the host that 
multiplexes the physical resources across multiple VMs. Each VM has its own OS layer (\ie guest OS). 
Unlike a VM, a container is built utilizing OS-level virtualization. Containers share a host's OS to access the underlying physical resources, instead of depending on the hypervisor.
The host's OS utilizes Linux namespaces and \textit{cgroups} to provide isolation between containers and restrict their access to system resources. 
Sharing the host's OS makes containers more lightweight. They can be provisioned more quickly  
compared to VMs~\cite{fn-startup-lanman21}. 

\noindent\textbf{Virtual switch (vSwitch):} vSwitches can be broadly classified into kernel-based approaches (\eg in-kernel Open vSwitch and Linux bridge) and userspace approaches that bypass the kernel (\eg OVS-DPDK~\cite{ovs-dpdk}, and OVS-AF\_XDP~\cite{ovs-afxdp-sigcomm}).
The kernel-based vSwitch runs within the host's OS kernel, using an in-kernel NIC driver to exchange packets with the physical NIC. The userspace vSwitch runs in the userspace of the host, using a userspace NIC driver to exchange packets with the physical NIC.

The userspace vSwitch relies on kernel-bypass to exchange packets with the NIC. We consider two distinct, but widely adopted, kernel-bypass architectures: DPDK~\cite{dpdk} and AF\_XDP~\cite{afxdp}. They both support zero-copy packet I/O between the NIC and userspace.
However, they are fundamentally different in the way they are driven to execute. DPDK's kernel-bypass depends only on polling while the kernel-bypass in AF\_XDP can be either event-driven (\ie triggered by each arriving packet) or polling.
DPDK implements a Poll Mode Driver (PMD), polling the NIC for received packets and packet transmission completions. This facilitates high-performance packet I/O between the NIC and the userspace functions. However, this leads to high CPU usage even if there is no incoming packet. An additional, specialized kernel driver (\eg UIO driver or VFIO driver) is required to block interrupt signals from the NIC, which helps the userspace PMD to work properly through active polling. However, this requires the NIC to be dedicated to DPDK. The exclusivity of DPDK leads to  compatibility problems between DPDK and the kernel stack; \eg the kernel stack now cannot access the NIC once DPDK has bound its kernel driver to the NIC.
One solution is to use Single Root I/O Virtualization (SR-IOV~\cite{Dong2010SRIOV}) to create multiple virtual Ethernet interfaces (these are called Virtual Functions, or VFs), and to dedicate DPDK's kernel driver to one of the VFs without disturbing the kernel stack (see \S\ref{sec:unified}).

AF\_XDP~\cite{afxdp}, is another kernel-bypass alternative to DPDK. The event-driven mode of AF\_XDP makes it strict load-proportional. Event-driven AF\_XDP executes only when a new packet arrives, thus it consumes \textit{no} CPU cycle when there is no packet. This fundamentally makes event-driven  AF\_XDP more resource-efficient under light load compared to DPDK. 
The polling mode AF\_XDP acts in a similar manner as DPDK. However, the polling mode of AF\_XDP still introduces interrupt overhead due to the execution of the XDP program at the NIC driver, which results in lower performance compared to DPDK. We evaluate both polling-based and event-driven AF\_XDP in \S\ref{sec:l2l3-setup}.
In addition, AF\_XDP (either polling or event-driven mode) does not require a specialized kernel driver to enable kernel-bypass, and thus it can  work seamlessly with the kernel stack to support protocol processing for an L4/L7 middlebox. DPDK on the other hand  requires SR-IOV support, in addition, to share the physical NIC with the kernel stack. Compared to a purely kernel-based solution (\ie using the kernel stack for both L2/L3 NFs and L4/L7 middleboxes), AF\_XDP achieves comparatively higher performance with zero-copy packet I/O between the NIC and userspace functions.

\noindent\textbf{Network protocol stack:}
The protocol stack can be kernel-based or could be in userspace, using kernel-bypass for passing packets. The kernel-based network protocol stack (\eg Linux kernel protocol stack) provides a full-function, robust, and proven solution for protocol processing, often with better usability than userspace protocol stack solutions such as Microboxes~\cite{Liu2018Microboxes} and mTCP~\cite{mtcp}, which provide limited support (\eg only TCP), thus limiting their usage. \textit{We primarily focus on the kernel-based protocol stack in this work.}

\noindent\textbf{Virtual device interfaces:} Typical virtual device interfaces include \textit{TUN/TAP}, \textit{veth pairs}, and \textit{virtio/vhost} devices.
\textit{TUN/TAP} operates as a data pipe (TUN for sending over L3 Tunnels, TAP for receiving L2 frames) that connects the kernel stack with userspace applications.
\textit{TUN/TAP} can work with \textit{virtio/vhost} virtual device interfaces to connect VMs or containers to the kernel-based vSwitch (Fig.~\ref{fig:ovs-datapath} (a) - (c)).
The \textit{virtio/vhost} interfaces execute as virtual NICs (vNICs) for VMs and containers.
The \textit{virtio} interface is in the VM/container, while the \textit{vhost} interface is in the host as the backend of the \textit{virtio} device. 
It is important to note that each has a userspace variant (\textit{virtio-user}, \textit{vhost-user}) as well as a kernel-based variant (\textit{virtio-net}, \textit{vhost-net}).
The \textit{virtio} variants and \textit{vhost} variants can be freely combined, \eg \textit{virtio-user} can work with \textit{vhost-net} (Fig.~\ref{fig:ovs-datapath} (a), (b)); \textit{virtio-net} can work with \textit{vhost-user} (Fig.~\ref{fig:ovs-datapath} (g)), \etc because they all follow the \textit{vhost} protocol~\cite{virtio-vhost-net}, having a consistent messaging APIs to work with different variants.
\textit{Veth pairs} are often used in container networking~\cite{assessing}, working as data pipes between the container's network namespace and the host's network namespace.
Unlike \textit{virtio/vhost}, the \textit{veth pair} works only in the kernel. It does not have a userspace variant, so it does not work directly with the userspace vSwitch (see Fig.~\ref{fig:ovs-datapath} (h)).

\vspace{-1mm}\subsection{Usability analysis of data plane models}\vspace{-1mm}
Fig.~\ref{fig:ovs-datapath} shows different variants for data plane connectivity for L2/L3 NFs and L4/L7 middleboxes by combining different options for virtualization, vSwitch, and virtual device interfaces. 
L2/L3 NFs do not require protocol layer processing, since they only offer an L2/L3  switch's forwarding capability, as in a vSwitch. L4/L7 middleboxes additionally require protocol stack processing. We first qualitatively evaluate the \textbf{usability} of different data plane models for L2/L3 NFs and L4/L7 middleboxes in Fig.~\ref{fig:ovs-datapath}, depending on whether the data plane model has a protocol stack or not.

The data plane models in Fig.~\ref{fig:ovs-datapath} (a), (b), (e), (f) do not involve protocol layer processing and are suitable for L2/L3 NFs. The data plane models in Fig.~\ref{fig:ovs-datapath} (c), (d), (g), (h), are all equipped with the kernel protocol stack and are suitable for L4/L7 middleboxes. Although data plane models for an L4/L7 middlebox (Fig.~\ref{fig:ovs-datapath} (c), (d), (g), (h)) can also be used for an L2/L3 NF. The protocol processing however adds unnecessary overhead, as it is  not required. In addition, we can extend the L2/L3 NF data plane models to support L4/L7 middleboxes by adding a userspace protocol stack; however, this approach is not favored by us for two reasons: (1) we want to use a full-function kernel protocol stack, and (2) having a separate userspace protocol stack in each middlebox function again adds to the memory footprint.

The use of the \textit{virtio-user} interface helps an L2/L3 NF data plane to bypass protocol layer processing, acting as the vNIC driver in a VM/container's userspace, directly interacting with the userspace function. Depending on the vSwitch being used, the \textit{virtio-user} device cooperates with different backend \textit{vhost} devices to create a direct data pipe between the userspace function and the vSwitch (either kernel-based or in userspace) to exchange raw packets: the \textit{vhost-net} device is used to connect with the kernel-based vSwitch through the \textit{TUN/TAP} (Fig.~\ref{fig:ovs-datapath} (a), (b));  the \textit{vhost-user} device is used to connect with the userspace vSwitch (Fig.~\ref{fig:ovs-datapath} (e), (f)).

When using containers to virtualize L4/L7 middleboxes (Fig.~\ref{fig:ovs-datapath} (d), (h)), the key element to enable the network protocol stack is the \textit{veth pair}. The container-side veth connects to the protocol stack in the container's network namespace (implemented in the host's kernel), for necessary protocol processing.\footnote{Note: there is no L2/L3 processing in the container's network namespace. The reason is the container actually shares the same kernel with the host. As the L2/L3 processing is performed by the kernel-based vSwitch in the host's network namespace, packets enter into the protocol layer stack after being passed to the container's network namespace. Thus, no duplicate L2/L3 processing is performed inside the container. Each \textit{veth pair} is assigned a unique IP address, which is used for L2/L3 forwarding across different containers' network namespaces. Applications with a container namespace share the same IP address and are differentiated by L4 port numbers.}
The host-side veth connects to host's network namespace, so it can seamlessly work with the kernel-based vSwitch (d). However, if we have to work with a userspace vSwitch (h), the packet needs to be injected from the userspace to the container's network namespace for protocol processing. To achieve this goal, the userspace vSwitch is connected to the kernel via the \textit{virtio-user/vhost-net} and \textit{TUN/TAP} device interfaces. The \textit{TUN/TAP} interface is configured with a point-to-point link to the \textit{veth pair}, which helps avoid duplicate L2/L3 processing in host's network namespace. 

When using VMs to virtualize L4/L7 middlebox functions, the \textit{virtio-net} device interface is used to utilize the protocol stack in VM's kernel. The \textit{virtio-net} device operates as the in-kernel vNIC driver, interacting with the userspace function through VM's kernel stack. Just like the \textit{virtio-user} device interface, the \textit{virtio-net} interface can work with either a kernel-based vSwitch (Fig.~\ref{fig:ovs-datapath} (c)) or a userspace vSwitch (Fig.~\ref{fig:ovs-datapath} (g)) by cooperating with specific backend \textit{vhost} device interface.

\vspace{-1mm}\subsection{Auditing Overheads of data plane models}\label{sec:auditing}\vspace{-1mm}

The data plane models in Fig.~\ref{fig:ovs-datapath}, with their selection of elements (\ie vSwitch, virtualization framework, virtual device interfaces) in constructing the data plane, may result in different data plane performance. Through a careful auditing of the overhead, we seek to identify the optimal data plane model for L2/L3 NFs and L4/L7 middleboxes. For this, we focus on the data plane overhead with a function chain.

For both L2/L3 NFs and L4/L7 middleboxes, function chains are mediated by the vSwitch to route packets between functions to be processed in the order they are configured in the chain. Additional protocol processing is required for the L4/L7 middlebox case.  We only show the auditing results when using
DPDK as the kernel-bypass architecture for the userspace vSwitch in this auditing. 

    \begin{figure}[t]
    \centering
        \includegraphics[width=.9\columnwidth]{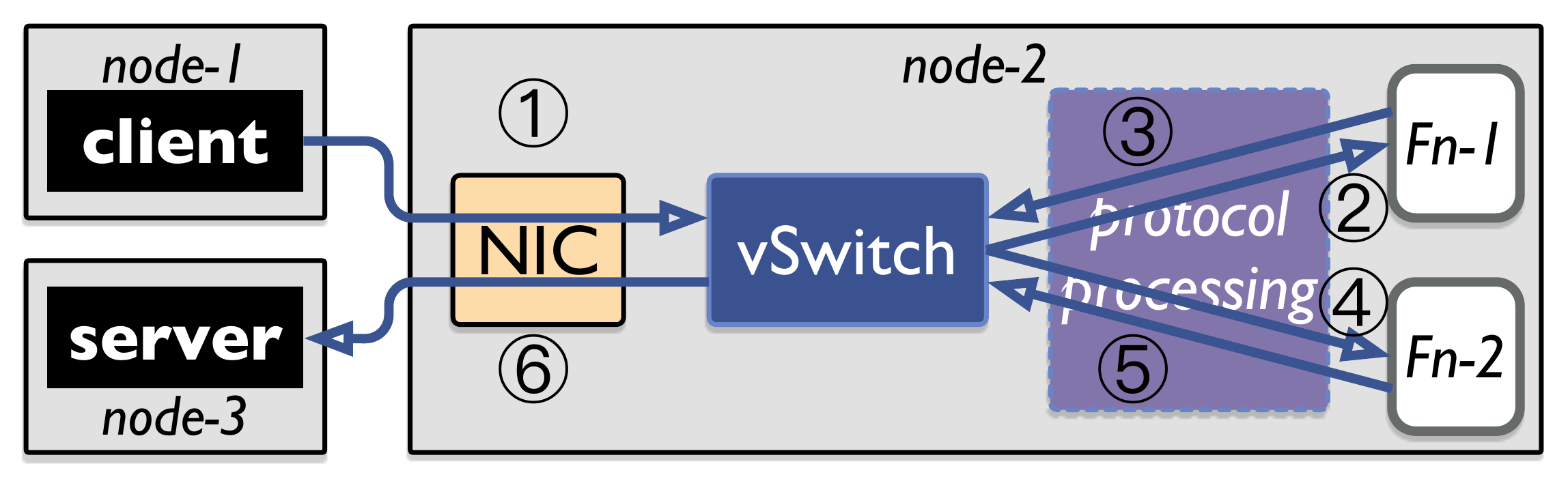}\vspace{-3mm}
    \caption{A generalized data pipeline for an NFV/Middlebox chain. Note: we only show the client-to-server datapath; protocol processing is only available for L4/L7 middlebox.}
    \vspace{-6mm}
    \label{fig:data-pipeline}
    \end{figure}

We use the abstract function chain setup of two functions (Fig.~\ref{fig:data-pipeline}) to 
represent the data pipeline for all cases. We assume functions in the same chain are placed on the same node so that there is no cross-node data transfer. The client sends  packets to the backend server through an intermediate node (node-2 in Fig.~\ref{fig:data-pipeline}) that implements the function chain. 
(\ding{172}) A packet first arrives at the physical NIC and is then passed to the vSwitch. (\ding{173}) The vSwitch routes the packet to the first function in the chain (Fn-1). (\ding{174}) After the first function completes processing the packet, the packet is sent back to the vSwitch. (\ding{175}) The vSwitch routes the packet to the next function in the chain (Fn-2). (\ding{176}) The second function processes the packet and returns it to the vSwitch. (\ding{177}) The vSwitch then routes the packet out through the NIC to the backend server.

\begin{table}[b]
\centering\vspace{-5mm}
\caption{Overhead auditing of L2/L3 NF data plane models}\vspace{-2mm}
\label{tab:l2l3}
\resizebox{\columnwidth}{!}{%
\begin{tabular}{|ccc|cc|cccc|c|}
\hline
\multicolumn{3}{|c|}{\multirow{2}{*}{\bf Data pipeline No.}} &
  \multicolumn{2}{c|}{\begin{tabular}[c]{@{}c@{}}Outside the chain\\ (NIC-vSwitch)\end{tabular}} &
  \multicolumn{4}{c|}{\begin{tabular}[c]{@{}c@{}}Within the chain \\ (Fn-vSwitch-Fn)\end{tabular}} &
  \multirow{2}{*}{\bf total} \\ \cline{4-9}
\multicolumn{3}{|c|}{} &
  \multicolumn{1}{c|}{\quad\ding{172}\quad\quad} &
  \ding{177} &
  \multicolumn{1}{c|}{\ding{173}} &
  \multicolumn{1}{c|}{\ding{174}} &
  \multicolumn{1}{c|}{\ding{175}} &
  \ding{176} &
   \\ \hline
\multicolumn{1}{|c|}{\multirow{4}{*}{\bf \# of copies}} &
  \multicolumn{1}{c|}{\multirow{2}{*}{\begin{tabular}[c]{@{}c@{}}kernel-based\\ vSwitch\end{tabular}}} &
  (a) &
  \multicolumn{1}{c|}{0} &
  0 &
  \multicolumn{1}{c|}{1} &
  \multicolumn{1}{c|}{1} &
  \multicolumn{1}{c|}{1} &
  1 &
  4 \\ \cline{3-10} 
\multicolumn{1}{|c|}{} &
  \multicolumn{1}{c|}{} &
  (b) &
  \multicolumn{1}{c|}{0} &
  0 &
  \multicolumn{1}{c|}{1} &
  \multicolumn{1}{c|}{1} &
  \multicolumn{1}{c|}{1} &
  1 &
  4 \\ \cline{2-10} 
\multicolumn{1}{|c|}{} &
  \multicolumn{1}{c|}{\multirow{2}{*}{\begin{tabular}[c]{@{}c@{}}userspace\\ vSwitch\end{tabular}}} &
  (e) &
  \multicolumn{1}{c|}{0} &
  0 &
  \multicolumn{1}{c|}{1} &
  \multicolumn{1}{c|}{1} &
  \multicolumn{1}{c|}{1} &
  1 &
  4 \\ \cline{3-10} 
\multicolumn{1}{|c|}{} &
  \multicolumn{1}{c|}{} &
  (f) &
  \multicolumn{1}{c|}{0} &
  0 &
  \multicolumn{1}{c|}{1} &
  \multicolumn{1}{c|}{1} &
  \multicolumn{1}{c|}{1} &
  1 &
  4 \\ \hline\hline
\multicolumn{1}{|c|}{\multirow{4}{*}{\bf  \# of interrupts}} &
  \multicolumn{1}{c|}{\multirow{2}{*}{\begin{tabular}[c]{@{}c@{}}kernel-based\\ vSwitch\end{tabular}}} &
  (a) &
  \multicolumn{1}{c|}{1} &
  0 &
  \multicolumn{1}{c|}{1} &
  \multicolumn{1}{c|}{1} &
  \multicolumn{1}{c|}{1} &
  1 &
  5 \\ \cline{3-10} 
\multicolumn{1}{|c|}{} &
  \multicolumn{1}{c|}{} &
  (b) &
  \multicolumn{1}{c|}{1} &
  0 &
  \multicolumn{1}{c|}{1} &
  \multicolumn{1}{c|}{1} &
  \multicolumn{1}{c|}{1} &
  1 &
  5 \\ \cline{2-10} 
\multicolumn{1}{|c|}{} &
  \multicolumn{1}{c|}{\multirow{2}{*}{\begin{tabular}[c]{@{}c@{}}userspace\\ vSwitch\end{tabular}}} &
  (e) &
  \multicolumn{1}{c|}{0} &
  0 &
  \multicolumn{1}{c|}{0} &
  \multicolumn{1}{c|}{0} &
  \multicolumn{1}{c|}{0} &
  0 &
  0 \\ \cline{3-10} 
\multicolumn{1}{|c|}{} &
  \multicolumn{1}{c|}{} &
  (f) &
  \multicolumn{1}{c|}{0} &
  0 &
  \multicolumn{1}{c|}{0} &
  \multicolumn{1}{c|}{0} &
  \multicolumn{1}{c|}{0} &
  0 &
  0 \\ \hline\hline
\multicolumn{1}{|c|}{\multirow{4}{*}{\begin{tabular}[c]{@{}c@{}}\bf  \# of context\\  switch\end{tabular}}} &
  \multicolumn{1}{c|}{\multirow{2}{*}{\begin{tabular}[c]{@{}c@{}}kernel-based\\ vSwitch\end{tabular}}} &
  (a) &
  \multicolumn{1}{c|}{0} &
  0 &
  \multicolumn{1}{c|}{1} &
  \multicolumn{1}{c|}{1} &
  \multicolumn{1}{c|}{1} &
  1 &
  4 \\ \cline{3-10} 
\multicolumn{1}{|c|}{} &
  \multicolumn{1}{c|}{} &
  (b) &
  \multicolumn{1}{c|}{0} &
  0 &
  \multicolumn{1}{c|}{1} &
  \multicolumn{1}{c|}{1} &
  \multicolumn{1}{c|}{1} &
  1 &
  4 \\ \cline{2-10} 
\multicolumn{1}{|c|}{} &
  \multicolumn{1}{c|}{\multirow{2}{*}{\begin{tabular}[c]{@{}c@{}}userspace\\ vSwitch\end{tabular}}} &
  (e) &
  \multicolumn{1}{c|}{0} &
  0 &
  \multicolumn{1}{c|}{0} &
  \multicolumn{1}{c|}{0} &
  \multicolumn{1}{c|}{0} &
  0 &
  0 \\ \cline{3-10} 
\multicolumn{1}{|c|}{} &
  \multicolumn{1}{c|}{} &
  (f) &
  \multicolumn{1}{c|}{0} &
  0 &
  \multicolumn{1}{c|}{0} &
  \multicolumn{1}{c|}{0} &
  \multicolumn{1}{c|}{0} &
  0 &
  0 \\ \hline
\end{tabular}%
}
\vspace{.5ex}

{\raggedright
(a) kernel-based vSwitch + virtio-user/vhost-net \& TUN/TAP + VM; \\
(b) kernel-based vSwitch + virtio-user/vhost-net \& TUN/TAP + container; \\
(e) userspace vSwitch + virtio-user/vhost-user + VM; \\
(f) userspace vSwitch + virtio-user/vhost-user + container;\par}

\vspace{.5ex}
{\raggedright\rightskip=0pt plus .2\hsize
\textbf{Note:} Context switches may happen when two userspace processes (\eg the NF and the vSwitch) are placed on the same CPU core. However, in NFV scenario, NFs and the vSwitch are typically dedicated with a separate CPU core, owing to the need of high performance. We assume NFs and the vSwitch assigned with dedicated CPU core in the overhead auditing. \textit{virtio-user} uses DPDK's PMD to send/receive packets. There is no interrupt involved.\par}
\end{table}

Table~\ref{tab:l2l3} shows the overhead auditing for the 
L2/L3 scenarios (Fig.~\ref{fig:ovs-datapath} (a), (b), (e), (f)). Table~\ref{tab:l4l7} shows the overhead auditing for the L4/L7 scenarios (Fig.~\ref{fig:ovs-datapath} (c), (d), (g), (h)). We do not include the switching/routing overhead (\ie cycles spent on forwarding/routing table lookup), as it is a necessary operation to exchange packets between functions (either L2/L3 or L4/L7) and cannot be avoided. We have several key takeaways below drawn from our  auditing of the packet flow.

\noindent\textbf{Takeaway\#1:} 
\textit{Using the \textbf{userspace vSwitch} in conjunction with  \textbf{virtio-user/vhost-user} ((e) and (f)) saves a significant amount of overhead, and is preferred for L2/L3 NFs.}

The userspace vSwitch does not show a significant overhead difference compared to the kernel-based vSwitch 
when moving the packet between the vSwitch and the NIC (\ding{172} and \ding{177}, see ``Outside the chain'' column in Table~\ref{tab:l2l3}).
Compared to the userspace vSwitch (using DPDK for kernel-bypass), the kernel-based vSwitch incurs one additional interrupt when receiving packets from the NIC. 

The advantage of the userspace vSwitch is the ability to work with userspace virtual device interfaces, \ie \textit{virtio-user/vhost-user}. 
Working in conjunction with \textit{virtio-user/vhost-user}, the userspace vSwitch does not incur an interrupt or context switch when passing packets within the function chain (\ding{173} to \ding{176}). On the other hand, the kernel-based vSwitch has to exchange the packet with the function in userspace through \textit{virtio-user/vhost-net} \& \textit{TUN/TAP} ((a) and (b)), which incurs an interrupt and a context switch each time the packet crosses the kernel-userspace boundary (\ding{173} to \ding{176}), a less desirable option. 
However, none of them avoid the data copies incurred when transmitting the packet within the chain (details below in \textbf{Takeaway\#3}).

\begin{table}[b]
\centering\vspace{-5mm}
\caption{Overhead auditing of L4/L7 middlebox data plane models}\vspace{-2mm}
\label{tab:l4l7}
\resizebox{\columnwidth}{!}{%
\begin{tabular}{|ccc|cc|cccc|c|}
\hline
\multicolumn{3}{|c|}{\multirow{2}{*}{\textbf{Data pipeline No.}}} &
  \multicolumn{2}{c|}{\begin{tabular}[c]{@{}c@{}}Outside the chain\\ (NIC-vSwitch)\end{tabular}} &
  \multicolumn{4}{c|}{\begin{tabular}[c]{@{}c@{}}Within the chain \\ (Fn-vSwitch-Fn)\end{tabular}} &
  \multirow{2}{*}{\textbf{total}} \\ \cline{4-9}
\multicolumn{3}{|c|}{} &
  \multicolumn{1}{c|}{\quad\ding{172}\quad\quad} &
  \ding{177} &
  \multicolumn{1}{c|}{\ding{173}} &
  \multicolumn{1}{c|}{\ding{174}} &
  \multicolumn{1}{c|}{\ding{175}} &
  \ding{176} &
   \\ \hline
\multicolumn{1}{|c|}{\multirow{4}{*}{\textbf{\# of copies}}} &
  \multicolumn{1}{c|}{\multirow{2}{*}{\begin{tabular}[c]{@{}c@{}}kernel-based\\ vSwitch\end{tabular}}} &
  (c) &
  \multicolumn{1}{c|}{0} &
  0 &
  \multicolumn{1}{c|}{2} &
  \multicolumn{1}{c|}{2} &
  \multicolumn{1}{c|}{2} &
  2 &
  8 \\ \cline{3-10} 
\multicolumn{1}{|c|}{} &
  \multicolumn{1}{c|}{} &
  (d) &
  \multicolumn{1}{c|}{0} &
  0 &
  \multicolumn{1}{c|}{1} &
  \multicolumn{1}{c|}{1} &
  \multicolumn{1}{c|}{1} &
  1 &
  4 \\ \cline{2-10} 
\multicolumn{1}{|c|}{} &
  \multicolumn{1}{c|}{\multirow{2}{*}{\begin{tabular}[c]{@{}c@{}}userspace\\ vSwitch\end{tabular}}} &
  (g) &
  \multicolumn{1}{c|}{0} &
  0 &
  \multicolumn{1}{c|}{2} &
  \multicolumn{1}{c|}{2} &
  \multicolumn{1}{c|}{2} &
  2 &
  8 \\ \cline{3-10} 
\multicolumn{1}{|c|}{} &
  \multicolumn{1}{c|}{} &
  (h) &
  \multicolumn{1}{c|}{0} &
  0 &
  \multicolumn{1}{c|}{2} &
  \multicolumn{1}{c|}{2} &
  \multicolumn{1}{c|}{2} &
  2 &
  8 \\ \hline\hline
\multicolumn{1}{|c|}{\multirow{4}{*}{\textbf{\# of interrupts}}} &
  \multicolumn{1}{c|}{\multirow{2}{*}{\begin{tabular}[c]{@{}c@{}}kernel-based\\ vSwitch\end{tabular}}} &
  (c) &
  \multicolumn{1}{c|}{1} &
  0 &
  \multicolumn{1}{c|}{2} &
  \multicolumn{1}{c|}{2} &
  \multicolumn{1}{c|}{2} &
  2 &
  9 \\ \cline{3-10} 
\multicolumn{1}{|c|}{} &
  \multicolumn{1}{c|}{} &
  (d) &
  \multicolumn{1}{c|}{1} &
  0 &
  \multicolumn{1}{c|}{2} &
  \multicolumn{1}{c|}{2} &
  \multicolumn{1}{c|}{2} &
  2 &
  9 \\ \cline{2-10} 
\multicolumn{1}{|c|}{} &
  \multicolumn{1}{c|}{\multirow{2}{*}{\begin{tabular}[c]{@{}c@{}}userspace\\ vSwitch\end{tabular}}} &
  (g) &
  \multicolumn{1}{c|}{0} &
  0 &
  \multicolumn{1}{c|}{2} &
  \multicolumn{1}{c|}{2} &
  \multicolumn{1}{c|}{2} &
  2 &
  8 \\ \cline{3-10} 
\multicolumn{1}{|c|}{} &
  \multicolumn{1}{c|}{} &
  (h) &
  \multicolumn{1}{c|}{0} &
  0 &
  \multicolumn{1}{c|}{3} &
  \multicolumn{1}{c|}{3} &
  \multicolumn{1}{c|}{3} &
  3 &
  12 \\ \hline\hline
\multicolumn{1}{|c|}{\multirow{4}{*}{\textbf{\# of context switch}}} &
  \multicolumn{1}{c|}{\multirow{2}{*}{\begin{tabular}[c]{@{}c@{}}kernel-based\\ vSwitch\end{tabular}}} &
  (c) &
  \multicolumn{1}{c|}{0} &
  0 &
  \multicolumn{1}{c|}{2} &
  \multicolumn{1}{c|}{2} &
  \multicolumn{1}{c|}{2} &
  2 &
  8 \\ \cline{3-10} 
\multicolumn{1}{|c|}{} &
  \multicolumn{1}{c|}{} &
  (d) &
  \multicolumn{1}{c|}{0} &
  0 &
  \multicolumn{1}{c|}{1} &
  \multicolumn{1}{c|}{1} &
  \multicolumn{1}{c|}{1} &
  1 &
  4 \\ \cline{2-10} 
\multicolumn{1}{|c|}{} &
  \multicolumn{1}{c|}{\multirow{2}{*}{\begin{tabular}[c]{@{}c@{}}userspace\\ vSwitch\end{tabular}}} &
  (g) &
  \multicolumn{1}{c|}{0} &
  0 &
  \multicolumn{1}{c|}{1} &
  \multicolumn{1}{c|}{1} &
  \multicolumn{1}{c|}{1} &
  1 &
  4 \\ \cline{3-10} 
\multicolumn{1}{|c|}{} &
  \multicolumn{1}{c|}{} &
  (h) &
  \multicolumn{1}{c|}{0} &
  0 &
  \multicolumn{1}{c|}{2} &
  \multicolumn{1}{c|}{2} &
  \multicolumn{1}{c|}{2} &
  2 &
  8 \\ \hline\hline
\multicolumn{1}{|c|}{\multirow{4}{*}{\textbf{\begin{tabular}[c]{@{}c@{}}\# of protocol \\ processing tasks\end{tabular}}}} &
  \multicolumn{1}{c|}{\multirow{2}{*}{\begin{tabular}[c]{@{}c@{}}kernel-based\\ vSwitch\end{tabular}}} &
  (c) &
  \multicolumn{1}{c|}{0} &
  0 &
  \multicolumn{1}{c|}{1} &
  \multicolumn{1}{c|}{1} &
  \multicolumn{1}{c|}{1} &
  1 &
  4 \\ \cline{3-10} 
\multicolumn{1}{|c|}{} &
  \multicolumn{1}{c|}{} &
  (d) &
  \multicolumn{1}{c|}{0} &
  0 &
  \multicolumn{1}{c|}{1} &
  \multicolumn{1}{c|}{1} &
  \multicolumn{1}{c|}{1} &
  1 &
  4 \\ \cline{2-10} 
\multicolumn{1}{|c|}{} &
  \multicolumn{1}{c|}{\multirow{2}{*}{\begin{tabular}[c]{@{}c@{}}userspace\\ vSwitch\end{tabular}}} &
  (g) &
  \multicolumn{1}{c|}{0} &
  0 &
  \multicolumn{1}{c|}{1} &
  \multicolumn{1}{c|}{1} &
  \multicolumn{1}{c|}{1} &
  1 &
  4 \\ \cline{3-10} 
\multicolumn{1}{|c|}{} &
  \multicolumn{1}{c|}{} &
  (h) &
  \multicolumn{1}{c|}{0} &
  0 &
  \multicolumn{1}{c|}{1} &
  \multicolumn{1}{c|}{1} &
  \multicolumn{1}{c|}{1} &
  1 &
  4 \\ \hline\hline
\multicolumn{1}{|c|}{\multirow{4}{*}{\textbf{\begin{tabular}[c]{@{}c@{}}\# of serialization or\\ deserialization (L7) \end{tabular}}}} &
  \multicolumn{1}{c|}{\multirow{2}{*}{\begin{tabular}[c]{@{}c@{}}kernel-based\\ vSwitch\end{tabular}}} &
  (c) &
  \multicolumn{1}{c|}{0} &
  0 &
  \multicolumn{1}{c|}{1} &
  \multicolumn{1}{c|}{1} &
  \multicolumn{1}{c|}{1} &
  1 &
  4 \\ \cline{3-10} 
\multicolumn{1}{|c|}{} &
  \multicolumn{1}{c|}{} &
  (d) &
  \multicolumn{1}{c|}{0} &
  0 &
  \multicolumn{1}{c|}{1} &
  \multicolumn{1}{c|}{1} &
  \multicolumn{1}{c|}{1} &
  1 &
  4 \\ \cline{2-10} 
\multicolumn{1}{|c|}{} &
  \multicolumn{1}{c|}{\multirow{2}{*}{\begin{tabular}[c]{@{}c@{}}userspace\\ vSwitch\end{tabular}}} &
  (g) &
  \multicolumn{1}{c|}{0} &
  0 &
  \multicolumn{1}{c|}{1} &
  \multicolumn{1}{c|}{1} &
  \multicolumn{1}{c|}{1} &
  1 &
  4 \\ \cline{3-10} 
\multicolumn{1}{|c|}{} &
  \multicolumn{1}{c|}{} &
  (h) &
  \multicolumn{1}{c|}{0} &
  0 &
  \multicolumn{1}{c|}{1} &
  \multicolumn{1}{c|}{1} &
  \multicolumn{1}{c|}{1} &
  1 &
  4 \\ \hline\hline
\multicolumn{1}{|c|}{\multirow{4}{*}{\textbf{\begin{tabular}[c]{@{}c@{}}\# of L2/L3\\ processing tasks\end{tabular}}}} &
  \multicolumn{1}{c|}{\multirow{2}{*}{\begin{tabular}[c]{@{}c@{}}kernel-based\\ vSwitch\end{tabular}}} &
  (c) &
  \multicolumn{1}{c|}{0} &
  1 &
  \multicolumn{1}{c|}{2} &
  \multicolumn{1}{c|}{1} &
  \multicolumn{1}{c|}{2} &
  1 &
  7 \\ \cline{3-10} 
\multicolumn{1}{|c|}{} &
  \multicolumn{1}{c|}{} &
  (d) &
  \multicolumn{1}{c|}{0} &
  1 &
  \multicolumn{1}{c|}{1} &
  \multicolumn{1}{c|}{0} &
  \multicolumn{1}{c|}{1} &
  0 &
  3 \\ \cline{2-10} 
\multicolumn{1}{|c|}{} &
  \multicolumn{1}{c|}{\multirow{2}{*}{\begin{tabular}[c]{@{}c@{}}userspace\\ vSwitch\end{tabular}}} &
  (g) &
  \multicolumn{1}{c|}{0} &
  1 &
  \multicolumn{1}{c|}{2} &
  \multicolumn{1}{c|}{1} &
  \multicolumn{1}{c|}{2} &
  1 &
  7 \\ \cline{3-10} 
\multicolumn{1}{|c|}{} &
  \multicolumn{1}{c|}{} &
  (h) &
  \multicolumn{1}{c|}{0} &
  1 &
  \multicolumn{1}{c|}{1} &
  \multicolumn{1}{c|}{0} &
  \multicolumn{1}{c|}{1} &
  0 &
  3 \\ \hline
\end{tabular}%
}
\vspace{.5ex}

{\raggedright
(c) kernel-based vSwitch + \textit{virtio-net/vhost-net} \& \textit{TUN/TAP} + VM; \\
(d) kernel-based vSwitch + \textit{veth} + container; \\
(g) userspace vSwitch + \textit{virtio-net/vhost-user} + VM; \\
(h) userspace vSwitch + \textit{virtio-user/vhost-net} \& \textit{TUN/TAP} + \textit{veth} + container\par}
\end{table}

\noindent\textbf{Takeaway\#2:} 
\textit{Using the \textbf{kernel-based vSwitch} in conjunction with  \textbf{veth} and container (d) incurs the least  overhead for L4/L7 middleboxes.}

Just as with the L2/L3 NF use case, the use of different vSwitches in L4/L7 middlebox case to exchange packets between the NIC and middlebox (\ding{172} and \ding{177}) does not have a significant difference. However, as L4/L7 middleboxes require kernel protocol processing, the kernel-based vSwitch has an advantage, as it can work
seamlessly with the protocol stack in the host's kernel. Since containers share the host's kernel, it is ideal to follow the data plane model (d) and connect the kernel-based vSwitch with the container via the \textit{veth pair}. As shown in Table~\ref{tab:l4l7}, each time when the packet is exchanged between the middlebox and the vSwitch (\ding{173} to \ding{176}), (d) it saves 1 data copy and 1 context switch compared to (c), which also adopts the kernel-based vSwitch. As (c) uses \textit{virtio-net/vhost-net} \& \textit{TUN/TAP} to connect VM and host's kernel, there is 1 data copy and 1 context switch involved.

The use of a userspace vSwitch along with the virtio-user/vhost-net interface (h) is also less preferable than (d). (h) with the userspace vSwitch differs from (d) (which uses the kernel-based vSwitch) because packets have to be looped back from the vSwitch in userspace to the kernel for protocol processing. This incurs one more data copy, interrupt, and context switch compared to (d), as seen in Table~\ref{tab:l4l7},
resulting in poorer performance.

Using the userspace vSwitch and the \textit{vhost-user} interface to work with a VM (g) is slightly better, as both the userspace vSwitch and the \textit{vhost-user} interface work in the userspace, thus eliminating one context switch compared to using the \textit{virtio-net/vhost-net} \& \textit{TUN/TAP} in (c). However, (g) still incurs an additional data copy because of the kernel-userspace boundary crossing within the VM. 
Moreover, as the packet has to traverse the entire VM's kernel stack in (c) and (g), there is unnecessary, duplicate L2/L3 processing involved in the VM's kernel in addition to the L2/L3 processing performed by the vSwitch in the host. This duplicate processing is avoided in (d) with the use of containers, which reuses the OS kernel from the host and avoids duplicate processing.

\noindent\textbf{Takeaway\#3:} 
\textit{Heavyweight service function chain for L2/L3 NFs and L4/L7 middleboxes.}

As shown in Table~\ref{tab:l2l3} and~\ref{tab:l4l7}, the major source of data plane overhead comes within the function chain (\ding{173} to \ding{176}). Even with the best combination we identified for L2/L3 NFs (f) and L4/L7 middleboxes (d), there are excessive data copies within a service function chain with existing solutions. With the best L2/L3 solution (f), one data copy is incurred each time a packet is passed from the vSwitch to the NF (\ding{173}, \ding{175}), and vice versa (\ding{174}, \ding{176}).
This also holds true for the best L4/L7 solution (d). The situation is worse for the L4/L7 case, as there are many additional overheads, including  interrupts, context switches, protocol processing tasks, and serialization/deserialization tasks, that are incurred for the communication within the chain (\ding{173} to \ding{176}).

\noindent\textbf{Discussion:}
Containers share the host's kernel protocol stack, resulting in a smaller memory footprint than having a dedicated kernel stack in each VM. This becomes important with scale, 
as the number of NFs/middleboxes grows. 
The smaller footprint contributes to faster startup of containerized functions~\cite{fn-startup-lanman21}. 
Containers also avoid duplicate L2/L3 processing for L4/L7 middleboxes (see \textbf{Takeaway\#2}).
For L2/L3 NFs, there is no significant difference in the data plane cost between VMs and containers (compare (e) and (f) in Table~\ref{tab:l2l3}). While we choose to work with containers, the design of \name is also generally applicable to a VM-based environment.

Data plane models (f) ``userspace vSwitch + \textit{virtio-user/vhost-user} + container'' and (d) ``kernel-based vSwitch + \textit{veth} + container'' are the best solution for L2/L3 NFs and L4/L7 middleboxes, respectively, as they introduce the minimal amount of overhead and are most lightweight against other alternatives. 
However, even the optimal data plane models are too heavyweight to construct
the function chain for L2/L3 NFs and L4/L7 middleboxes.
In fact, the overhead in the current service function chain design builds as the size of the function chain increases, which can result in significant performance loss. Unnecessary packet processing overhead is introduced in the data transfer between vSwitch and functions, as well as expensive protocol processing (for L4/L7 only). All these factors make it difficult for us to achieve a high-performance NFV/middlebox framework.

\vspace{-1mm}\section{Shared memory communication in \name}\label{sec:shm-design}\vspace{-1mm}

    \begin{figure}[b]
    \vspace{-5mm}
    \centering
        \includegraphics[width=.9\columnwidth]{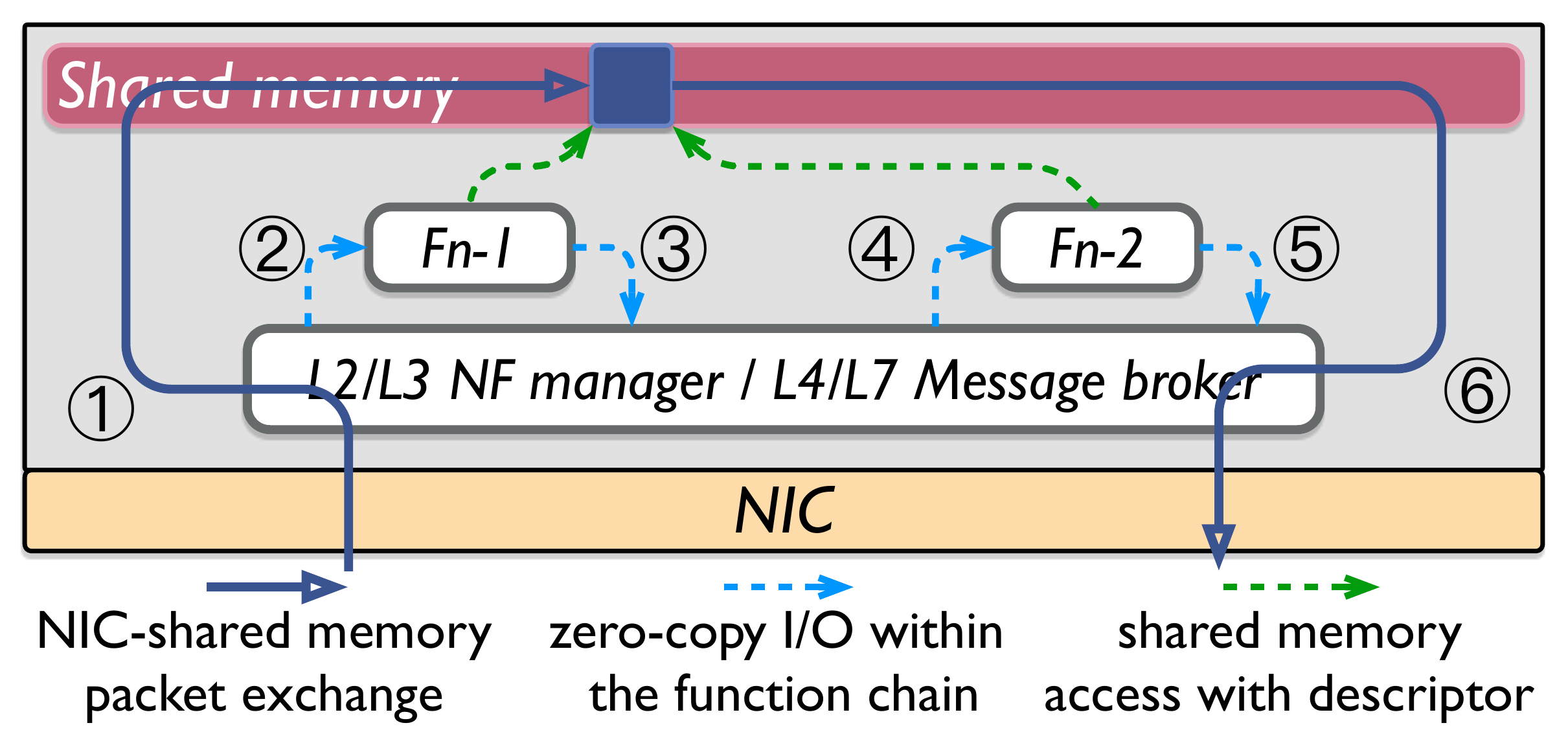}\vspace{-5mm}
    \caption{A generalized shared memory communication data pipeline for a function chain in \name. Note: we only show the client-to-server datapath}
    \label{fig:shm-pipeline}
    \end{figure}

Shared memory communication can alleviate the data movement overheads of the
data plane within a function chain by keeping the data in a userspace memory pool to be shared by different functions in the chain. 
Fig.~\ref{fig:shm-pipeline} shows a generalized data pipeline using shared memory communication in \name. It is a chain, with two functions (either L2/L3 NFs or L4/L7 middlebox functions), both on the same host.
Steps \ding{172} and \ding{177} move the packets between the NIC and shared memory, while \ding{173} to \ding{176} pass packet descriptors between functions to achieve zero-copy packet delivery within the function chain. 
An intermediate component (running in userspace) is used to provide forwarding/routing support within the function chain, which is similar to the vSwtich in Fig.~\ref{fig:ovs-datapath}. We call this intermediate component the ``NF manager'' in the L2/L3 scenario, or ``message broker'' in the L4/L7 scenario.
The NF manager/message broker is responsible for moving packets between the NIC and the shared memory in steps \ding{172} and \ding{177}.

Three key elements enable shared memory communication for a function chain: 
(1) \textbf{NIC-shared memory packet exchange}. An incoming packet is moved into the userspace shared memory prior to processing by the function chain (either L2/L3 NF chain or L4/L7 middlebox chain); (2) \textbf{Zero-copy I/O within the function chain}. Instead of moving the data from one function to another, shared memory communication achieves zero-copy I/O within the function chain, by passing a pointer, which is the packet descriptor, to the data in shared memory. This substantially reduces overhead;
(3) \textbf{Shared memory support}. A memory pool is initialized and mapped to each function in the chain before it can be accessed.
There are multiple alternatives, with significant differences, for the ``NIC-shared memory packet exchange'' and ``zero-copy I/O within the function chain'' operations, which we now describe \textit{qualitatively}.
\subsubsection{NIC-shared memory packet exchange}\label{sec:nic-shm}
There are two distinct options: one approach bypasses the kernel, the other is a kernel-based approach. The kernel-bypass approach DMA's the packet to shared memory without involving the kernel stack. \textit{Exploiting kernel-bypass avoids heavyweight kernel processing and is better suited for building L2/L3 NFs as a `bump-in-the-wire'.}
As discussed in \S\ref{sec:basic}, the kernel-bypass approach can be further classified into a polling-based kernel-bypass (\ie with DPDK's PMD) and event-driven kernel-bypass (\ie using AF\_XDP). The NF manager (Fig.~\ref{fig:shm-pipeline}) works with these kernel-bypass alternatives to move packets between the NIC and shared memory (details in \S\ref{sec:dpdk-l2l3} and \S\ref{sec:ebpf-l2l3}).

The kernel-based approach, on the other hand, uses the kernel stack to pass packets between the NIC and the message broker in the userspace. The message broker exchanges packets with the kernel stack via the Linux socket interface. It then moves packets to shared memory for zero-copy processing within the function chain. This inevitably introduces overheads (\eg copy, context switch, etc) when a packet crosses the kernel-userspace boundary. It also incurs the overhead of kernel protocol layer processing, which is only useful for L4/L7 middleboxes.
\textit{The kernel-based approach is ideal for L4/L7 middleboxes, as it provides necessary processing using a full-function kernel protocol stack.}

\subsubsection{Zero-copy I/O for function chaining}\label{sec:zero-copy-chain}
Zero-copy I/O for function chaining can also be broadly implemented using either: (1) polling-based zero-copy I/O, \eg DPDK's RTE RING~\cite{dpdk-ring_lib}; or (2) event-driven zero-copy I/O, \eg eBPF's \skmsg~\cite{skmsg}.
It's important to understand the difference between these two options and their impact on  performance. 

eBPF's \skmsg is a socket-related eBPF program type, ``\texttt{BPF\_PROG\_TYPE\_SK\_MSG}''~\cite{skmsg}. \skmsg is attached to the socket of the function during its creation. It processes packets sent/received on the attached socket to/from the kernel.  The execution of \skmsg is triggered by the arrival of a packet, which is strictly event-driven and is thus load-proportional. Working in conjunction with the eBPF socket map (\texttt{BPF\_MAP\_TYPE\_SOCKMAP}~\cite{sockmap}), which provides necessary routing information, \skmsg can deliver packet descriptors between functions. The other option, DPDK's RTE RING, is implemented as a circular FIFO queue, used for buffering packet descriptors. Dedicated for each function is a Receive (RX) and Transmit (TX) ring pair to pass packet descriptors using polling.\footnote{Note: Polling the RTE ring does not require the simultaneous use of DPDK's PMD. It can be simply implemented as a \texttt{while} loop.} 
A function polls its own RX ring (using \texttt{rte\_ring\_dequeue()}) to receive packet descriptors and enqueue packet descriptors to its TX ring (using \texttt{rte\_ring\_enqueue()}) for transmission. A centralized routing component on the other side polls the TX ring of each function and moves queued packet descriptors to the RX ring of the destination function, based on its internal routing table.

\subsubsection{Shared memory support}\label{sec:shm-support}
\name uses DPDK's multi-process support~\cite{dpdk-multi_proc_support} to construct shared memory between functions within a service chain. We utilize a shared memory manager (running as a DPDK primary process\footnote{The DPDK primary process has privileges, enabling it to initialize memory pools in huge pages.}) to manage shared memory pools. 
During the initialization stage of \name, the shared memory manager in \name creates a private memory pool, with a unique ``shared data file prefix'' specified to isolate with other shared memory pools on the same node.
The ``shared data file prefix'' is used by DPDK's EAL to create hugepage files (\ie actual file system objects for DPDK's memory pools) in the Linux file system. A DPDK process is allowed to access a hugepage file, only if the same file prefix was specified during its creation. Additional details are in Appendix~\ref{appendix:dpdk-shm-support}, including shared memory support for VM-based functions.
We leverage this feature to build a security domain for \name that enhances the security of using shared memory for communication between NFs (see \S\ref{sec:security-domain}).

Each key element described is independent of the other, \eg using DPDK's multi-process doesn't require DPDK's PMD. So using DPDK's multi-process support to manage memory sharing between different functions incurs no polling overhead.

\noindent\textbf{Overhead Auditing \& Discussion:}
We perform overhead auditing of the function chain using shared memory communication. We consider two distinct approaches for both the L2/L3 NFs and L4/L7 middleboxes use cases: the polling-based approach (using DPDK's PMD and RTE RING), and the event-driven approach (using eBPF's AF\_XDP and \skmsg).

To conserve space, we have summarized the main takeaways here. A detailed discussion can be found in Appendix~\ref{sec:shm-auditing}.
The overhead auditing clearly shows the advantage of using shared memory communication, to reduce the overhead in almost every dimension (\eg data copy, interrupt, context switch, etc).
Thus, we factor it into our NFV/middlebox framework, \name. It is clear that L2/L3 \name should consider \textit{kernel-bypass} NIC-shared memory packet exchange to facilitate high performance. L4/L7 \name adopts \textit{kernel-based} NIC-shared memory packet exchange to provide the needed protocol processing.
We understand the trade-off between a polling-based solution and an event-driven solution by implementing the alternatives, and evaluating their performance, to help us decide which to use for \name.

\vspace{-1mm}\section{Design of \name: L2/L3 NFV}\label{sec:l2l3-middlenet}\vspace{-1mm}
We discuss the eBPF-based  and DPDK-based alternatives for L2/L3 NFV support, given the performance requirement of operating at line rate and being capable of supporting service function chains. Since they operate at L2/L3, there is less emphasis on having a full-function protocol stack.

\vspace{-1mm}\subsection{Overview}\label{sec:l2l3-overview}\vspace{-1mm}

\noindent\textbf{NIC-userspace kernel-bypass:}
\name takes full advantage of zero-copy packet delivery and kernel-bypass to move packets between the NIC and the userspace shared memory, so as to minimize overheads, reduce resource consumption, and achieve full line-rate L2/L3 packet processing (\S\ref{sec:nic-shm}). We consider two kernel-bypass alternatives:
polling-based DPDK's PMD and event-driven AF\_XDP (\S\ref{sec:basic}).

\noindent\textbf{Zero-copy I/O for function chaining:}
We evaluate two alternatives for L2/L3 \name, the \textit{polling-based} approach and the \textit{event-driven} approach. 
The polling-based alternative adopts DPDK's PMD for NIC-to-userspace delivery using kernel-bypass and DPDK's RTE RING for function chaining. The event-driven alternative adopts AF\_XDP for NIC-to-userspace kernel-bypass and \skmsg for function chains. This helps us evaluate the trade-off between performance and resource efficiency when using a polling-based design or an event-driven design to achieve a `bump-in-the-wire' L2/L3 NFV environment. 
Both of them use DPDK's multi-process support to manage the shared memory of L2/L3 \name (\S\ref{sec:shm-support}).
We implement these two alternatives based on OpenNetVM's design~\cite{opennetvm}, that is similar in principle to the design described in Fig.~\ref{fig:shm-pipeline}, \S\ref{sec:shm-design}.

\vspace{-1mm}\subsection{The DPDK-based L2/L3 NFV design}\label{sec:dpdk-l2l3}\vspace{-1mm}
The DPDK-based approach can be `expensive' in having dedicated CPU cores for polling.
In addition to the NF manager that dedicates one CPU core for the PMD, for each NF of the L2/L3 function chain, one CPU core is used up for each function to poll its RTE RING.
This can be wasteful if incoming traffic is low. Somewhat more complex NFV support, such as NFVnice~\cite{Kulkarni2020NFVnice}, can be used to mitigate these overheads by sharing a CPU core across multiple NFs.

    \begin{figure}[t]
    \centering
        \includegraphics[width=\columnwidth]{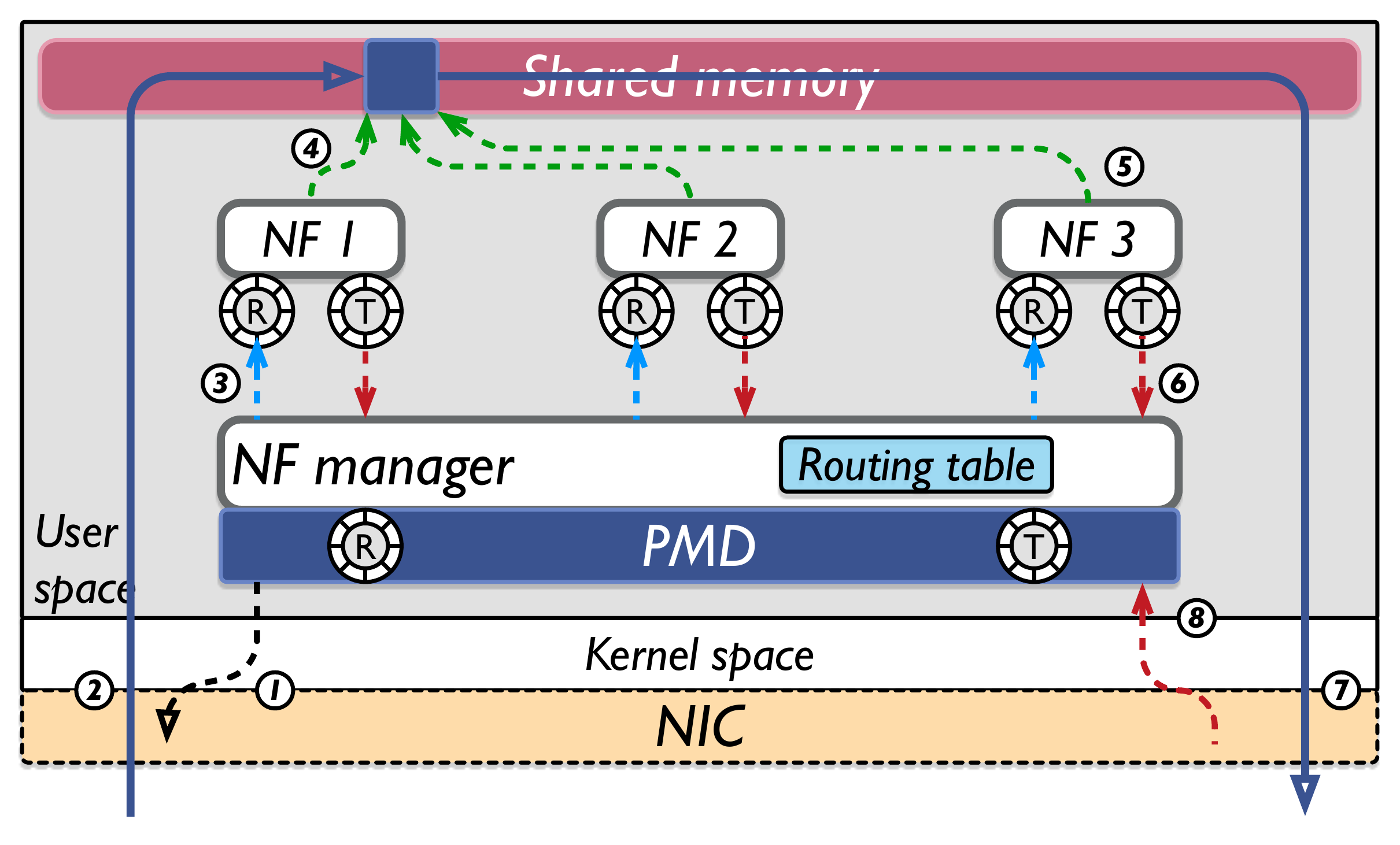}\vspace{-5mm}
    \caption{Packet processing flow for DPDK-based L2/L3 NFV: RX and TX}
    \label{fig:dpdk-nfv}\vspace{-5mm}
    \end{figure}

Fig.~\ref{fig:dpdk-nfv} depicts the packet flow of DPDK-based L2/L3 NFs. In the RX path, PMD provides a packet descriptor for the NIC (\ding{172}) to deliver the packet into the shared memory via DMA (\ding{173}). The NF manager examines the packet, and moves the packet descriptor into the RX ring of the target NF (\ding{174}), based on the routing table. The target NF obtains the packet descriptor by polling its RX ring and uses it to access the packet in shared memory  (\ding{175}). After the NF's packet processing is complete (\ding{176}), the NF writes the descriptor to its TX ring  (\ding{177}). On the other side, the NF manager continuously polls the NF's TX ring and sets up the packet transmission based on the descriptor in the ring (\ding{178}). The PMD then completes the processing once the packet is transmitted, to clean up the transmit descriptor  (\ding{179}).
Both TX and RX rings are polled by the PMD for RX and TX from/to the NIC, and NFs use polling to RX or TX packet descriptors. 

\noindent\textbf{Service function chains:} The NF manager utilizes destination information in the packet descriptor to support routing within an NF chain for the DPDK-based approach.
The routing table in the NF manager is used to resolve that NF's ID, thus
avoiding the need for each NF to maintain a private routing table. 
After the NF manager gets a packet descriptor from the TX ring of an NF,
it parses the packet descriptor to look at the destination NF information.
It then pushes a packet descriptor to the RX ring of the next NF to transfer ownership of the shared memory frame (as pointed to by the descriptor). Ownership for write is based on the NF currently owning a descriptor to that frame in shared memory, thus ensuring a single writer and obviating the need for locks. Using the NF manager for `centralized' routing mitigates contention when multiple NFs may forward to a downstream NF. 

\vspace{-1mm}\subsection{The eBPF-based L2/L3 NFV design}\label{sec:ebpf-l2l3}
\begin{figure}[t]
    \centering
        \includegraphics[width=\columnwidth]{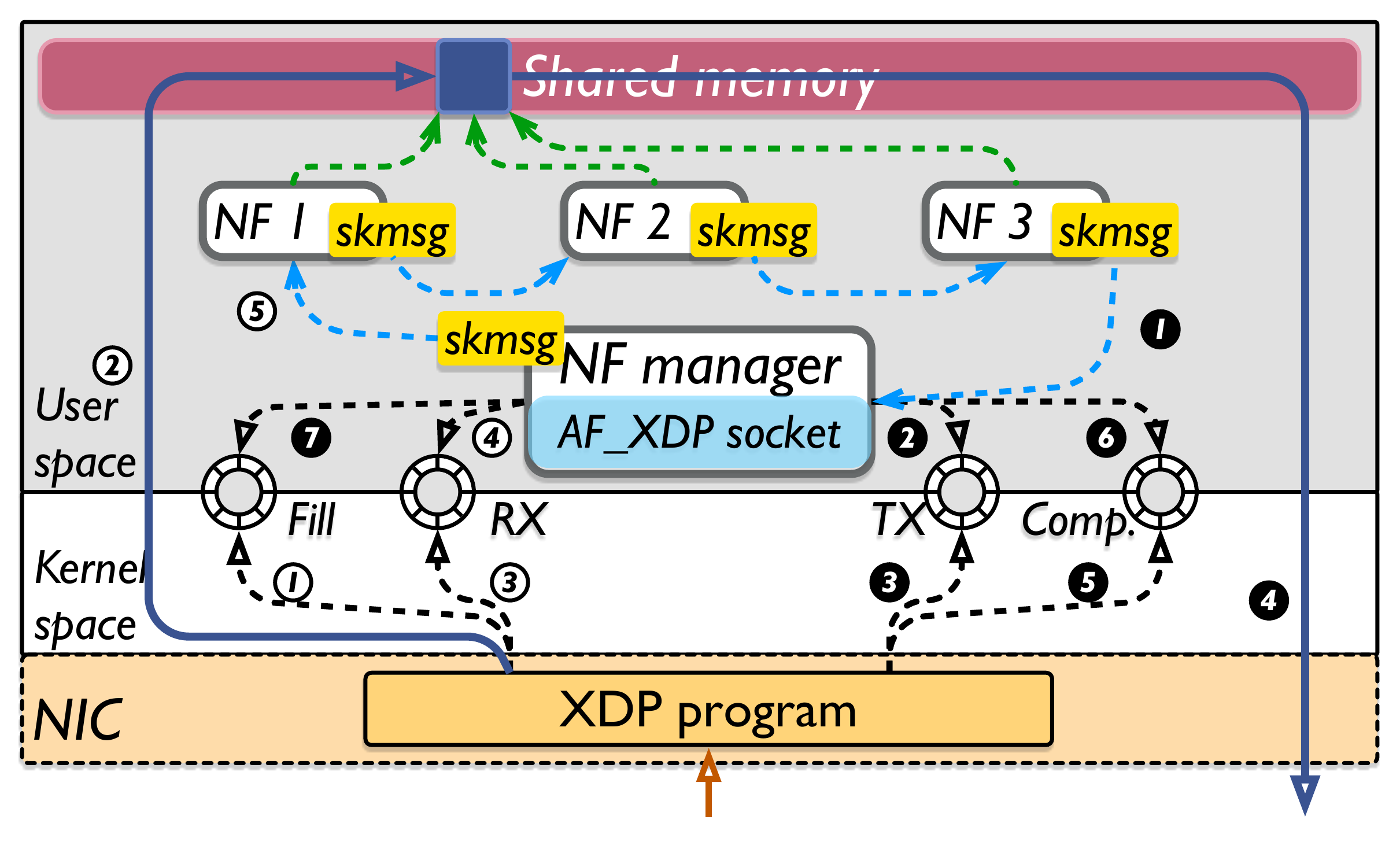}\vspace{-5mm}
    \caption{Packet processing flow for eBPF-based L2/L3 NFV: RX and TX}
    \label{fig:nfv}\vspace{-5mm}
    \end{figure}

\vspace{-1mm}
The NF manager in the eBPF-based L2/L3 \name opens
a dedicated AF\_XDP socket (\ie \xsk~\cite{afxdp}) that serves as an interface to interact with the kernel to handle RX and TX for AF\_XDP-based packet delivery. Each \xsk is assigned  a set of RX and TX rings to pass packet descriptors containing pointers to packets in shared memory. All \texttt{XSKs}
share a set of `Completion' and `Fill' rings, owned by the kernel and used to transfer ownership of the shared memory frame between the kernel and userspace NFs. AF\_XDP depends on interrupts triggered by the event execution of the XDP program attached to the NIC driver (Fig.~\ref{fig:nfv}). This interrupt notifies the packet processing component in userspace. However, these interrupts have to be managed with care  to avoid poor overload behavior when subjected to high packet rates~\cite{mogul1997eliminating}.

Fig.~\ref{fig:nfv} depicts the zero-copy packet flow based on AF\_XDP. 
An XDP program works in the kernel space with the NIC driver to handle packet reception (and transmission). The NIC is provided a descriptor (\ding{172}) pointing to an empty frame in shared memory. Upon reception, the packet is DMAed into shared memory (\ding{173}), and a receive interrupt  triggers an XDP\_REDIRECT which
moves the packet descriptor to the RX ring of the NF manager (\ding{174}) before invoking it.
In the interrupt service routine, the kernel notifies the NF manager about updates in its RX ring, which the NF manager then accesses via its \xsk (\ding{175}). The interrupt service routine is completed once the NF manager fetches the packet descriptor from the RX ring.
The NF manager invokes the corresponding NF (\ding{176}) and waits for NFs to complete processing.

    \begin{figure}[b]
    \vspace{-4mm}
    \centering
        \includegraphics[width=\columnwidth]{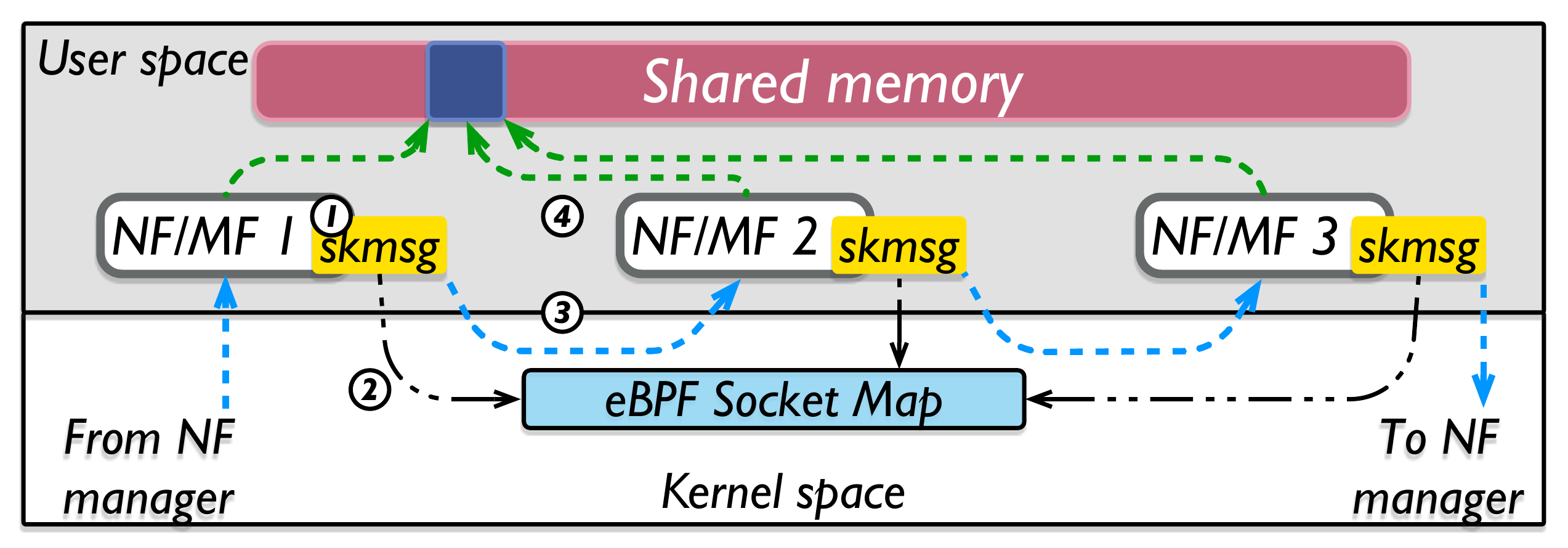}\vspace{-3mm}
    \caption{Function chaining in \name: eBPF-based approach}
    \label{fig:chain}
    \end{figure}
    
After the NF completes packet processing, the NF manager is invoked to transmit the packet out of the node (\ding{182}). The descriptor is populated in the TX ring (\ding{183}).
The system call by the NF manager (typically \texttt{sendmsg()}) notifies the kernel about the TX event (\ding{184}). The kernel then transmits the packet based on the descriptor given in the TX ring (\ding{185}). If the packet is successfully transmitted, the kernel pushes the descriptor back to the `Completion' ring (\ding{186}) to inform the NF manager that the frame can now be reused for the subsequent transmission.
The NF manager fetches the packet descriptor from the `Completion' ring  (\ding{187}) and moves it to the `Fill' ring for incoming packets  (\ding{188}).

We implement the NF manager with three threads to manage the different rings without locks. We use one thread to handle the read of the RX ring (\ding{175}) and another one to handle the transmit to the TX ring (\ding{183}). We use a third thread to coordinate between the `Completion' ring and the `Fill' ring. This thread watches for the kernel to move packet descriptors into the `Completion' ring (\ding{187}) upon transmitting completions. The third thread
then moves the packet descriptor from the  `Completion' ring to the `Fill' ring  (\ding{188}).

\begin{figure*}[htbp]
\centering
 \subfigure{\includegraphics[width=0.25\textwidth]{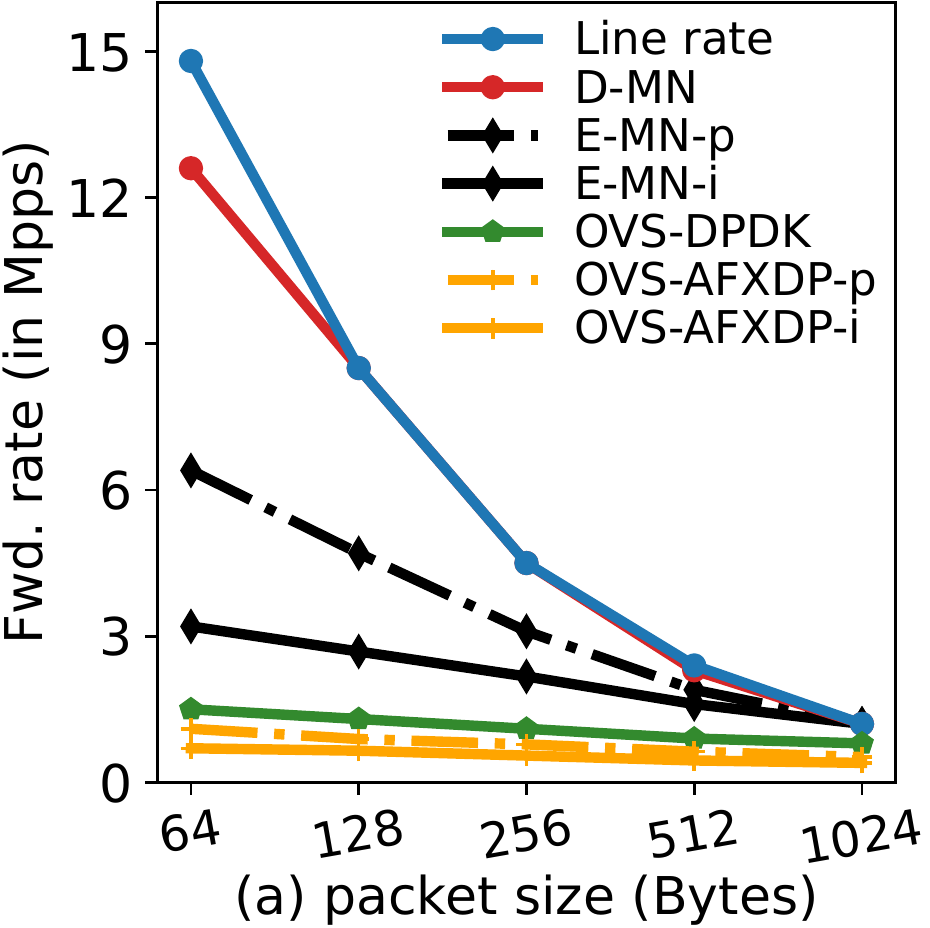}\label{fig:l2l3-pps}} \hfill
 \subfigure{\includegraphics[width=0.45\textwidth]{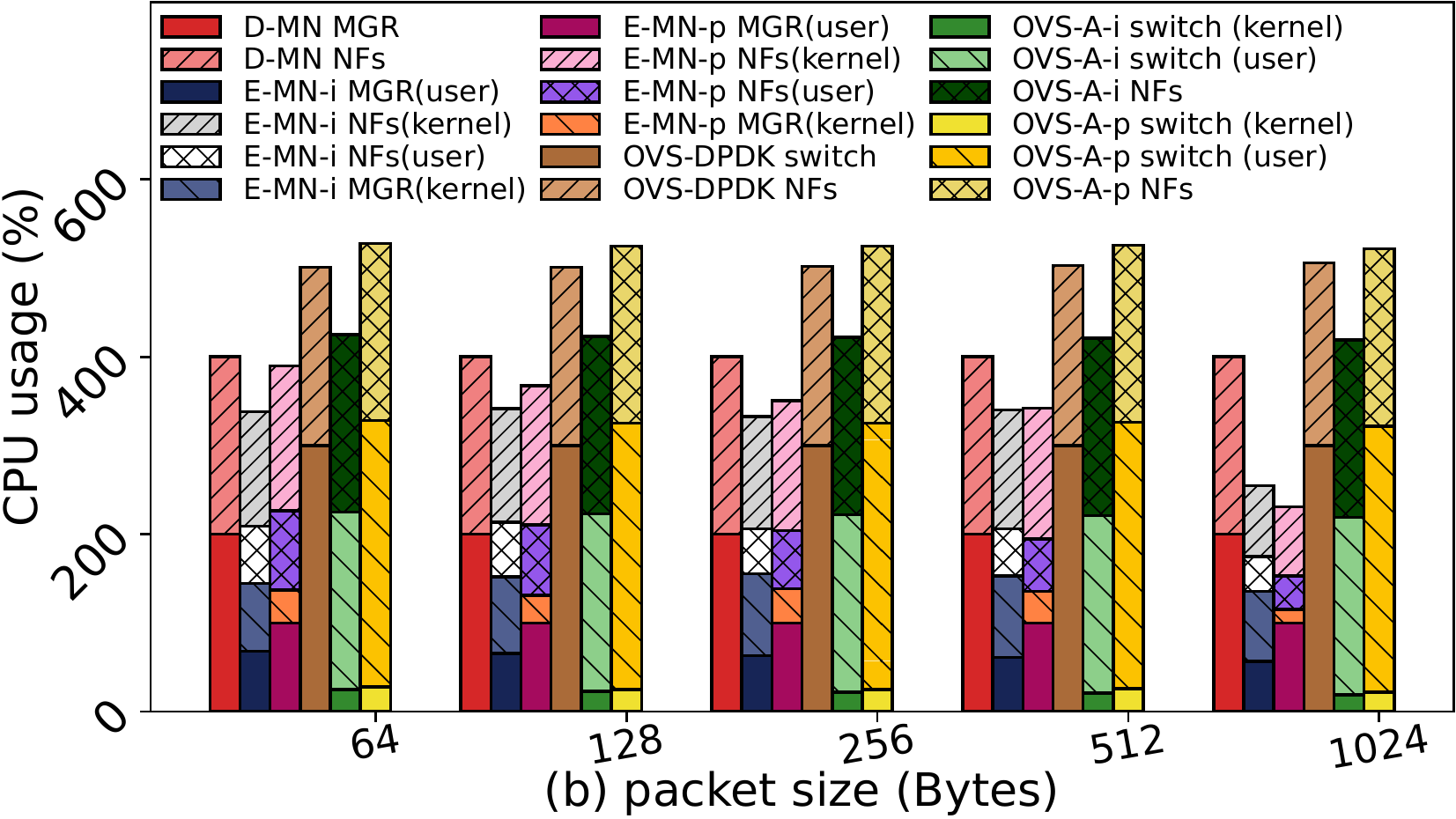}\label{fig:l2l3-cpu}} \hfill
 \subfigure{\includegraphics[width=0.25\textwidth]{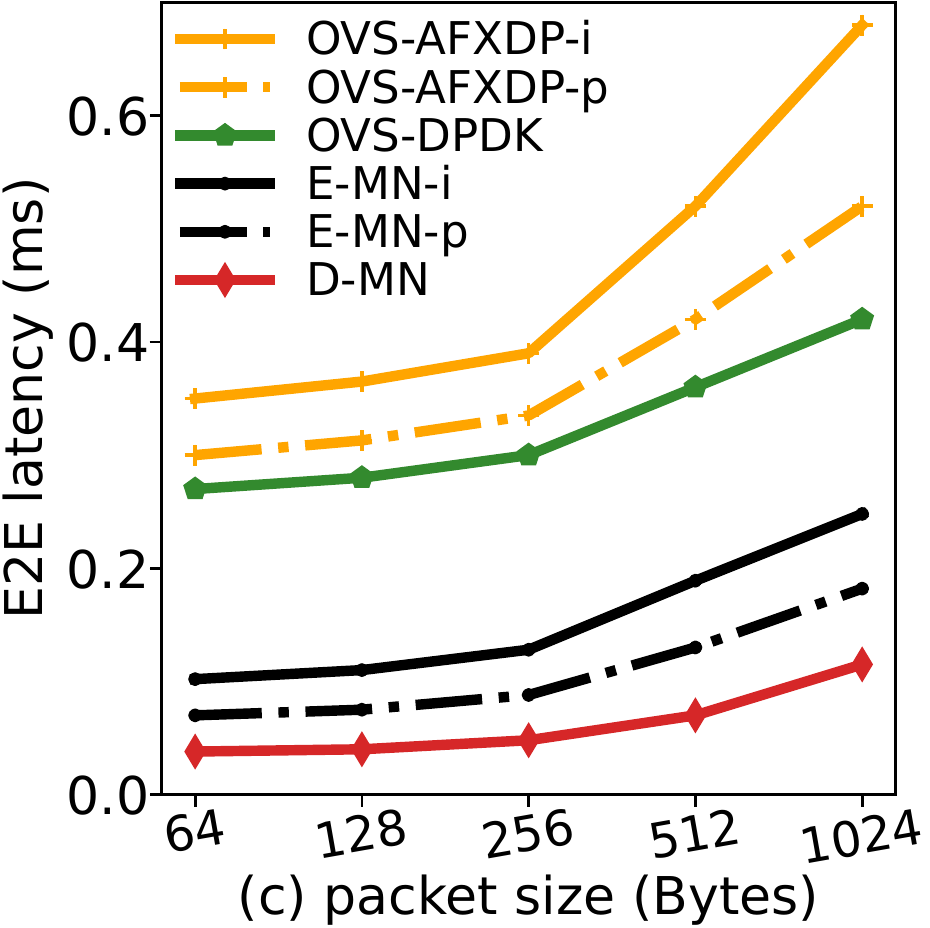}\label{fig:l2l3-latency}}\hfill
\vspace{-3mm}
\caption{Comparison between different L2/L3 alternatives: (a) Maximum loss free rate (MLFR) under different packet sizes, (b) CPU usage under MLFR under different packet sizes, (c) end-to-end latency under MLFR under different packet sizes. \textbf{Note:} \textit{D-MN} refers to \dpdk; \textit{E-MN-i} refers to \eBPF with interrupt-driven AF\_XDP socket; \textit{E-MN-p} refers to \eBPF with polling-based AF\_XDP socket; \textit{OVS-A-i} refers to OVS-AF\_XDP with interrupt-driven AF\_XDP socket; \textit{OVS-A-p} refers to OVS-AF\_XDP with polling-based AF\_XDP socket. }
\label{fig:l2l3-perf}\vspace{-5mm}
\end{figure*}

\noindent\textbf{Service function chains:} The eBPF-based L2/L3 approach uses \skmsg to support NF chains.
To support flexible routing between functions, we utilize eBPF's socket map. The in-kernel socket map maintains a map between the ID of the target NF and the socket interface information. As shown in Fig.~\ref{fig:chain}, the NF creates a packet descriptor to be sent (\ding{172}). The \skmsg performs a lookup in the socket map to determine the destination socket (\ding{173}). It then redirects the packet descriptor to the next NF (\ding{174}). That NF uses the descriptor to access  data in shared memory (\ding{175}) and passes the packet descriptor to the next NF  through \skmsg after processing.

\vspace{-1mm}\subsection{Performance evaluation}\label{sec:l2l3-setup}\vspace{-1mm}
\noindent\textbf{Experiment setup:}
We compare the performance of DPDK (\ie polling-based, hereafter referred to as \dpdk) and eBPF (\ie event-driven, hereafter referred to as \eBPF) approaches to support L2/L3 NFVs
with a `packet-centric' evaluation by comparing the Maximum Loss Free Rate (MLFR), the end-to-end latency, and CPU utilization at this MLFR for different packet sizes.
We use the data plane model (f) in \S\ref{sec:basic} as the primary baseline to compare with. For this, we choose two implementations of Open vSwitch as the kernel-bypass vSwitch in (f): OVS-DPDK~\cite{ovs-dpdk} and OVS-AF\_XDP~\cite{ovs-afxdp-sigcomm}. 
We set up our experiments on NSF Cloudlab~\cite{Dmitry2019CloudLab} with three nodes: the 1st node is configured with a \emph{Pktgen}~\cite{Pktgen} load generator for L2/L3 NFV use case; the 2nd node is configured with two \name alternatives (\dpdk, \eBPF) and the two OVS alternatives (OVS-DPDK, OVS-AF\_XDP).
The 3rd node is configured to return the packets directly back to the 1st node, to measure latency.
Each node has a 40-core CPU, 192GB memory, and a 10Gbps NIC. We use Ubuntu 20.04 with kernel version 5.15. We use DPDK v21.11~\cite{dpdk} and \emph{libbpf}~\cite{libbpf} v0.6.0 for eBPF-related experiments.

To achieve the best possible performance for OVS-DPDK and OVS-AF\_XDP baselines, we enable the ``Multiple Poll-Mode Driver Threads''~\cite{ovs-dpdk-multi-pmd} feature in OVS.
Each PMD thread runs on a dedicated CPU core and continually polls the physical NIC or the \textit{vhost-user} (Fig.~\ref{fig:ovs-datapath} (f)) to process incoming packets. OVS-AF\_XDP uses polling to retrieve packets from the NIC by default. For this polling-based OVS-AF\_XDP option (OVS-AF\_XDP-p, Fig.~\ref{fig:ovs-datapath} (f)), and OVS-DPDK, we create three PMD threads to achieve the highest performance.
We additionally configure the AF\_XDP socket in OVS-AF\_XDP  to run in the interrupt mode (\ie OVS-AF\_XDP-i)~\cite{afxdp-nonpmd}.\footnote{To enable the interrupt mode for AF\_XDP,  a user needs to specify the device type of the physical NIC as ``afxdp-nonpmd'' when attaching it to OVS.} This helps to move packets between NIC and userspace OVS in an event-driven manner. But, to achieve the optimal packet exchange performance between OVS-AF\_XDP-i and NFs, we use polling to avoid interrupt overheads for packet exchanges between OVS and the NFs. Only a data copy overhead is incurred between  OVS and the NFs when using polling on both sides. For this, we create two PMD threads to poll packets for getting packets to and from NFs (via \textit{vhost-user}).
For NFs in both the OVS-DPDK and OVS-AF\_XDP setups, each \textit{virtio-user} is dedicated with a CPU core to poll packets from OVS.
We also configure the AF\_XDP socket in \eBPF to operate in polling mode (\eBPF-p) and compare with the interrupt-based AF\_XDP socket (\eBPF-i).

We set up two NFs in a chain on the 2nd node: 
an L3 routing function followed by an L2 forwarding function. For  the L3 routing function, \name updates the IP address of received packets, and the L2 forwarding function of a subsequent NF in the chain updates the MAC address of received packets and forwards it to the 3rd node.
We collect the average value measured across 5 repetitions. Each run is for 60 seconds. 

\noindent\textbf{Discussion:}
Fig.~\ref{fig:l2l3-pps} shows the MLFR for different alternatives.
\dpdk achieves almost the line rate for different packet sizes. The exception is for packet sizes of 64Bytes, achieving 12.6M packets/sec (84\% of line rate) because of our limit on the number of CPU cores for the NF Manager and the PMD. Even with the limited CPU cores, \dpdk outperforms both \eBPF-i and \eBPF-p.
For a packet size of 64Bytes, \eBPF-i is limited to a forwarding rate of 3.2 Mpps (only 25\% of \dpdk) while \eBPF-p is limited to a forwarding rate of 6.3 Mpps (50\% of \dpdk). Moreover, if the NFs have more complex processing or if the load were to be higher (\eg if there is bidirectional traffic), then we observe receive-livelock~\cite{mogul1997eliminating}. The performance of \eBPF-i is limited by its overheads, including a number of interrupts and context switches (see Table~\ref{tab:l2l3-shm}).
As we observe in Fig.~\ref{fig:l2l3-cpu}, \eBPF-i's NF manager and the NFs themselves spent most of the CPU time in the kernel (53\% for the NF manager, 67\% for NFs) to handle interrupts generated by AF\_XDP socket or \skmsg, thus leaving fewer resources to perform the NF packet forwarding tasks.
\eBPF-p reduces interrupts by operating the AF\_XDP socket in polling mode, which helps it achieve better throughput compared to \eBPF-i. But, the performance of \eBPF-p is still worse than \dpdk as the execution of XDP program in the NIC driver is triggered by interrupts, in addition to the \skmsg overhead, all of which negatively impact the packet forwarding performance.
Although devoting more resources to \eBPF's NF manager and the NFs may alleviate this overload, it only postpones the problem when the traffic load continues to increase. Moreover, using more resources to mitigate overload defeats the original intention of using eBPF-based event-driven processing since the goal of using it is for resource efficiency. Focusing on the end-to-end packet latency, \dpdk achieves a 2.6$\times$ improvement compared to \eBPF-i, and is 1.8$\times$ better compared to \eBPF-p (Fig.~\ref{fig:l2l3-latency}).

Note that as the packet size increases, the CPU usage of both \eBPF-i and \eBPF-p is even lower compared to the other options. For example, at a packet size of 1024Bytes, the CPU usage of \eBPF-i and \eBPF-p are 63\% and 58\% of \dpdk, respectively. Since \eBPF-i and \eBPF-p use {\it event-driven} shared memory communication, as the packet size increases and the packet rate decreases (bounded by the line rate of the NIC used in this experiment). The overhead for \eBPF-i and \eBPF-p, which is strictly proportional to the packet rate, diminishes. 
Thus the CPU overhead reduces for larger packet sizes for \eBPF-i and \eBPF-p, which makes the event-driven design attractive for larger packet sizes for L2/L3 NFs. However, the event-driven approach still suffers from poor performance and relatively high CPU usage in handling L2/L3 traffic with smaller packet sizes. On the other hand, \dpdk maintains good performance across a range of packet sizes. Further, \dpdk can utilize the scheduling principles in NFVnice~\cite{Kulkarni2020NFVnice} to reduce the CPU consumption by multiplexing a CPU core across multiple NFs.

Both \dpdk and \eBPF outperform OVS-DPDK and OVS-AF\_XDP in terms of MLFR for receiving packets and latency. Looking at the CPU usage of OVS-DPDK, even though OVS-DPDK dedicates enough CPU resources (3 CPU cores for the OVS switch, one CPU core per NF) to achieve the best performance, the forwarding rate for it is worse than \eBPF.
This shows the negative impact of excessive data copies within the chain (\S\ref{sec:auditing}). Even though \eBPF also incurs interrupts and context switches (Table~\ref{tab:l4l7-shm}) in the data pipeline, as shown in Fig.~\ref{fig:shm-pipeline}, 
its exploitation of shared memory communication fundamentally improves the data plane performance of function chains, as discussed in Appendix~\ref{sec:shm-auditing}.
OVS-AF\_XDP on the other hand performs poorly. Running OVS-AF\_XDP in polling mode (OVS-AF\_XDP-p) improves throughput and reduces latency compared to running OVS-AF\_XDP in interrupt mode. This is because OVS-AF\_XDP-i suffers the overhead of interrupts and context switches for moving packets between the NIC and userspace, just like \eBPF-i. But the improvement of OVS-AF\_XDP-p is limited, particularly because of the data copy overhead within the chain.

\dpdk does constantly consume considerable CPU (one CPU core per NF, 2 CPU cores for the NF manager).
While this is a concern, its superior performance makes it more attractive for L2/L3 NFs, since they have to act like a `bump-in-the-wire'.
\eBPF is less attractive because of its poor overload behavior.

\vspace{-1mm}\section{Design of \name: L4/L7 Middlebox}\label{sec:l4l7-middlenet}\vspace{-1mm}

We discuss the corresponding eBPF-based and DPDK-based designs to support L4/L7 middleboxes. Since an L4/L7 middlebox relies heavily on protocol processing, we discuss optimizations, leveraging the kernel protocol stack processing, focusing on resource efficiency.

    \begin{figure}[b]
    \vspace{-5mm}
    \centering
        \includegraphics[width=\columnwidth]{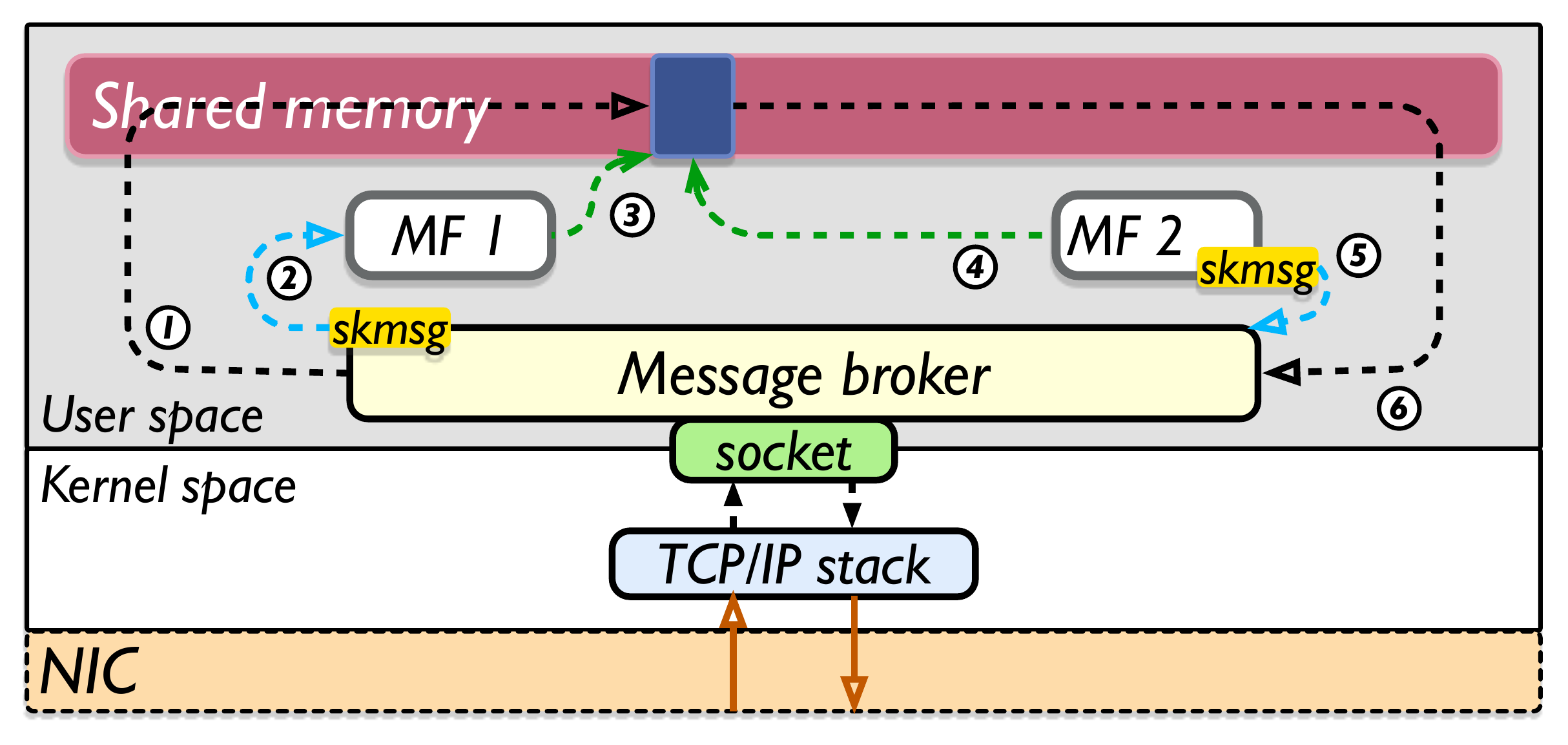}\vspace{-5mm}
    \caption{Packet processing flow for eBPF-based L4/L7 middleboxes}
    \label{fig:middleboxes}
    \end{figure}

\vspace{-1mm}\subsection{Overview}\vspace{-1mm}

\noindent\textbf{Protocol processing support:} Unlike L2/L3 NFs, packets pass through the kernel for the required protocol layer processing for L4/L7 middleboxes. L4/L7 \name uses a message broker (Fig.~\ref{fig:shm-pipeline}) to leverage the protocol processing 
in the kernel stack. Incoming packets processed by the kernel network protocol stack are delivered through a socket to a message broker in userspace.
This comes at a cost (see Appendix~\ref{sec:shm-auditing}), but \name benefits significantly from a fully functional in-kernel protocol stack
for L4/L7 middleboxes.

\noindent\textbf{Zero-copy I/O for function chaining \& shared memory support:}
We follow a similar methodology as in \S\ref{sec:l2l3-middlenet} to evaluate what is the most suited zero-copy I/O capability for function chains in L4/L7 \name.
For the eBPF-based L4/L7 middlebox design, packets are forwarded between MFs using eBPF's \skmsg capability.
For DPDK-based L4/L7 middlebox functionality, the message broker delivers descriptor entries to the ring of the target MF, with the payload in shared memory, after protocol processing by the message broker.

\vspace{-1mm}\subsection{The eBPF-based L4/L7 middlebox design}\label{sec:skmsg-l4l7}\vspace{-1mm}

Fig.~\ref{fig:middleboxes} depicts the packet flow for the eBPF-based L4/L7 \name. For inbound traffic, after the payload is moved into shared memory by the message broker (\ding{172}), a packet descriptor 
is sent to the target MF via \skmsg (\ding{173}). The MF then uses the descriptor to access the data in shared memory (\ding{174}). 
For outbound traffic, once the MF has finished processing the packet (\ding{175}), it uses \skmsg 
to inform the message broker (\ding{176}), which then fetches the packet in shared memory (\ding{177}) and transmits it on the network via the kernel protocol stack.

\noindent\textbf{Function chain support:}
The eBPF-based L4/L7 \name utilizes the eBPF's \skmsg and socket map for delivering packet descriptors within the function chain (similar to what we described for L2/L3 NFV with eBPF), as shown in Fig.~\ref{fig:chain}.
Although the eBPF-based L4/L7 approach still executes in a purely interrupt-driven manner, since the kernel protocol stack is involved, it often uses a flow-controlled transport protocol.
This potentially avoids overloading the receiver, and therefore, receive-livelocks are less of a concern. 
Interrupt-based processing does not use up a CPU like polling, so it is more resource-efficient and 
benefits the L4/L7 use case. We further mitigate the impact of interrupts with batching.

    \begin{figure}[b]
    \vspace{-5mm}
    \centering
        \includegraphics[width=\columnwidth]{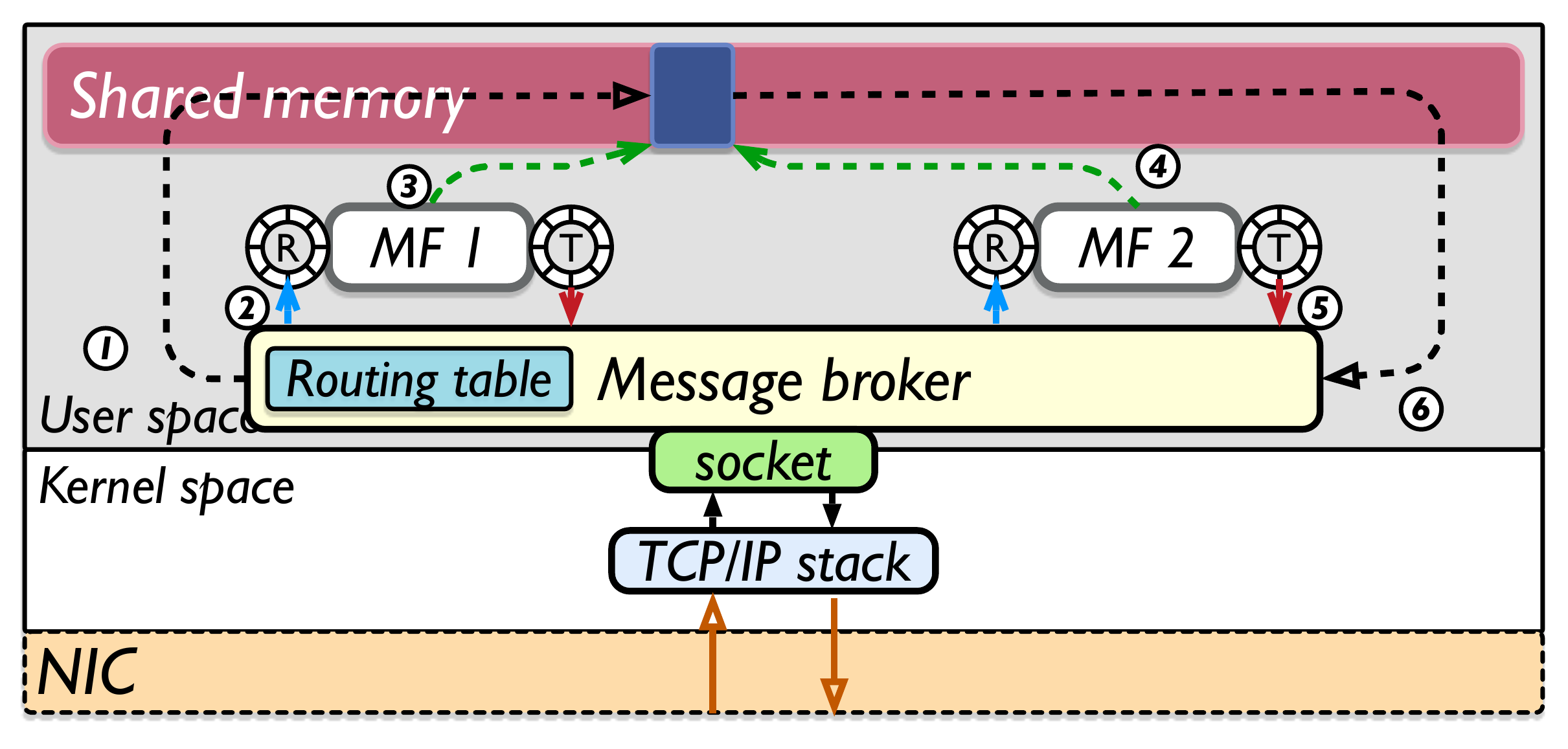}\vspace{-5mm}
    \caption{Packet processing flow for DPDK-based L4/L7 middleboxes}
    \label{fig:dpdk-middleboxes}
    \end{figure}

\noindent\textbf{Adaptive batching of \skmsg Processing:}
Since bursty traffic can cause a large number of \skmsg transfers,
we consider an adaptive batching mechanism to reduce the overhead of frequent \skmsg transfers. For each interrupt generated by \skmsg, instead of reading only one packet descriptor present in the socket buffer, we read multiple (up to a limit) packet descriptors available in the socket buffer. Thus, we can reduce the total number of interrupts, even for frequent \skmsg transfers, and mitigate overload behavior.

\vspace{-1mm}\subsection{The DPDK-based L4/L7 middlebox design}\vspace{-1mm}

To leverage the kernel protocol stack, we restructure the NF manager of the L2/L3 use case (Fig.~\ref{fig:dpdk-nfv}) into a message broker in the DPDK-based L4/L7 \name. 
The message broker writes the received payload to shared memory (\ding{172}), then, consulting the routing table, pushes the packet descriptor to the RX ring of the target MF (\ding{173}). The MF
keeps polling its RX ring for arriving packets. The MF uses the received packet descriptor to access the packet in shared memory and processes it (\ding{174}). Once the processing is complete (\ding{175}), the MF pushes the packet descriptor to its TX ring. 
On the other side, the message broker polls the TX ring of MFs for the packet descriptor (\ding{176}), then accesses the shared memory and sends the packet out through the kernel protocol stack (\ding{177}). 

\noindent\textbf{Function chain support:} 
The function chain support in the DPDK-based L4/L7 \name is the same as in the DPDK-based L2/L3 NFV use case (\S\ref{sec:dpdk-l2l3}). Here, the message broker 
performs the (same) tasks to transfer packet descriptors between MFs.

\begin{figure*}[htbp]
\centering
 \subfigure{\includegraphics[width=0.66\columnwidth]{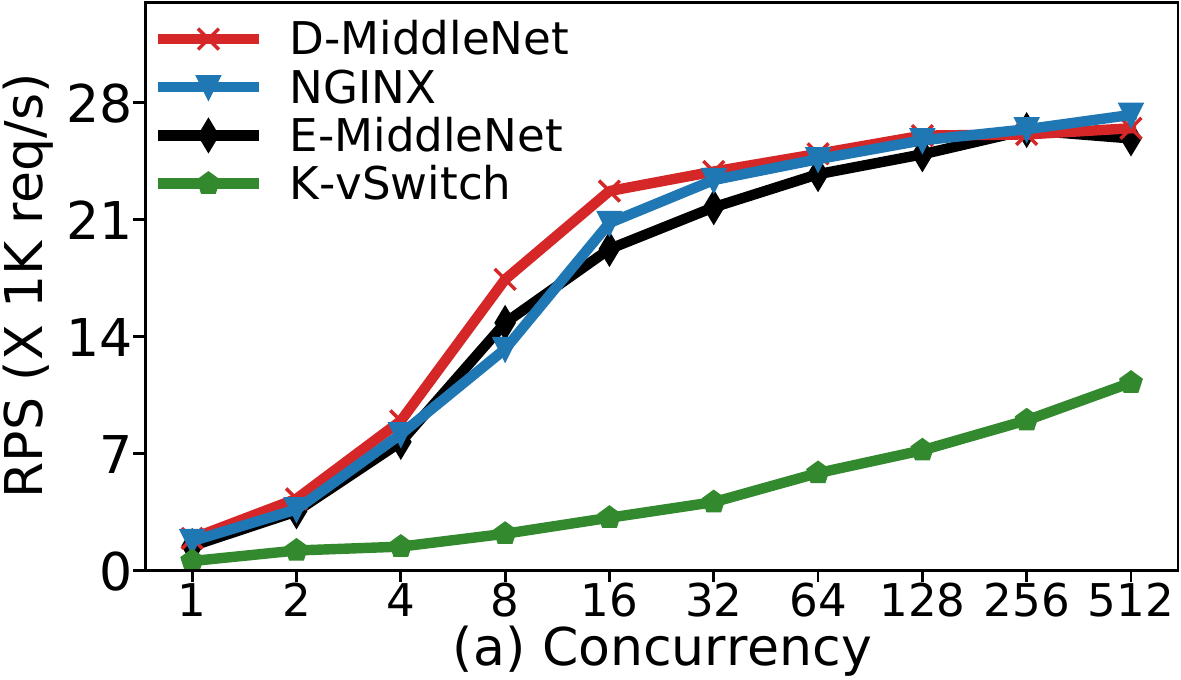}\label{fig:l4l7-rps}} \hfill
 \subfigure{\includegraphics[width=0.66\columnwidth]{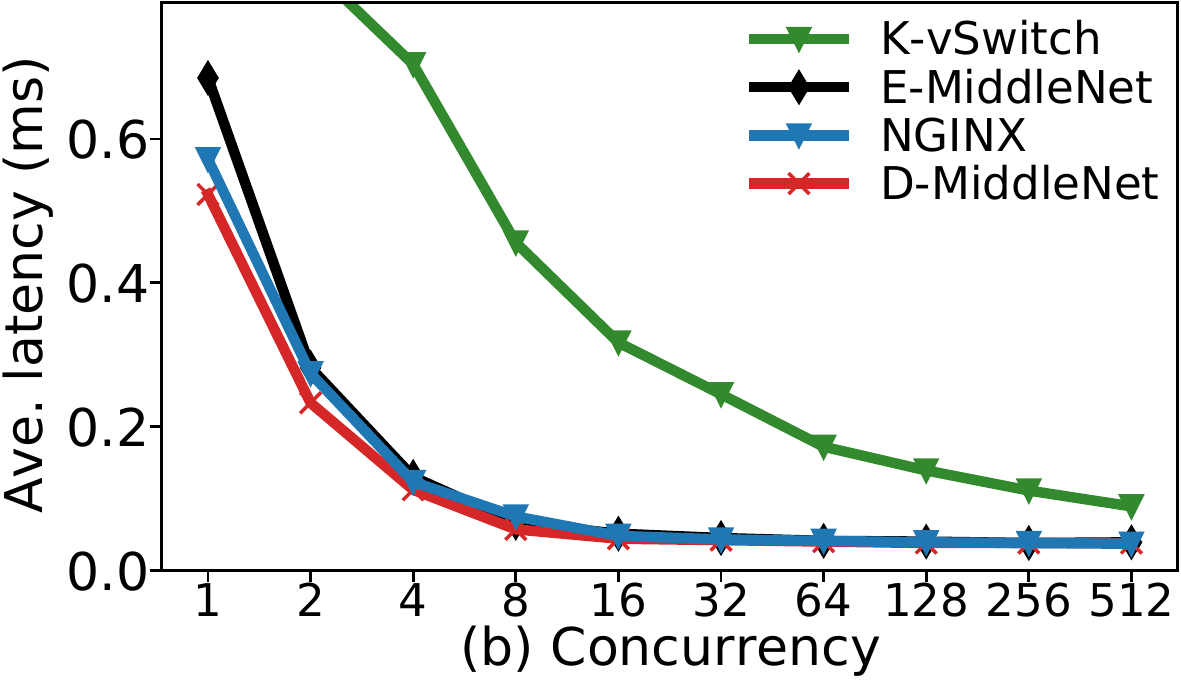}\label{fig:l4l7-latency}}\hfill
 \subfigure{\includegraphics[width=0.66\columnwidth]{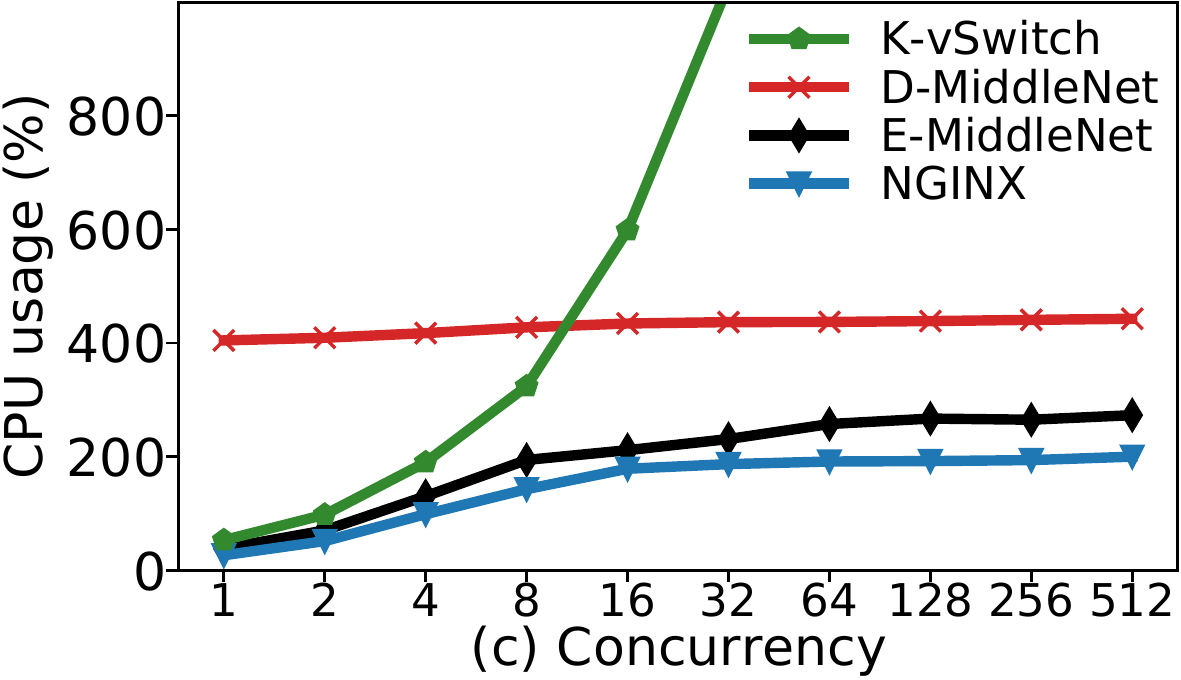}\label{fig:l4l7-cpu}}
\vspace{-3mm}
\caption{RPS (a), latency (b) and CPU usage (c) comparison between different L4/L7 middlebox approaches. Note: The CPU usage of the data plane model (d) exceeds 10 CPU cores at concurrency level 32 and consumes 30 CPU cores at concurrency level 512.}
\label{fig:l4l7-perf}\vspace{-5mm}
\end{figure*}

\begin{figure}[b]
\vspace{-5mm}
\centering
 \subfigure{\includegraphics[width=0.32\columnwidth]{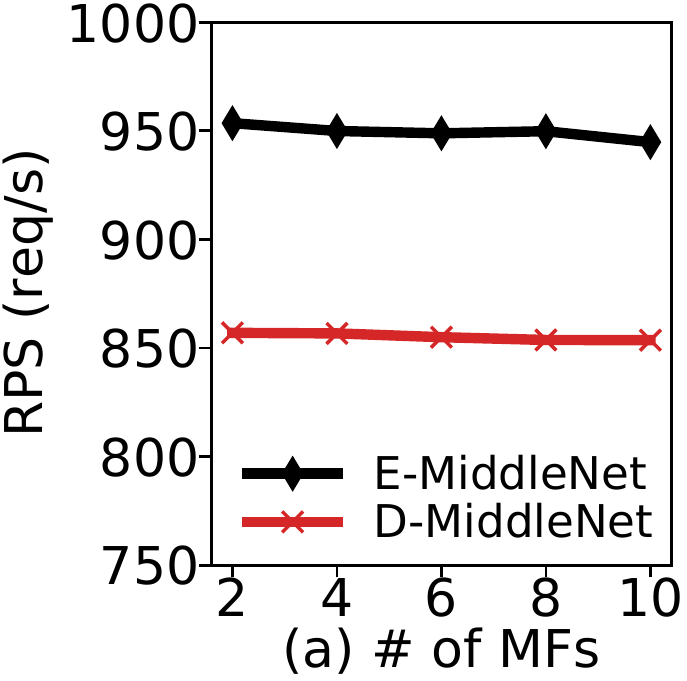}\label{fig:scale-rps}} \hfill
 \subfigure{\includegraphics[width=0.32\columnwidth]{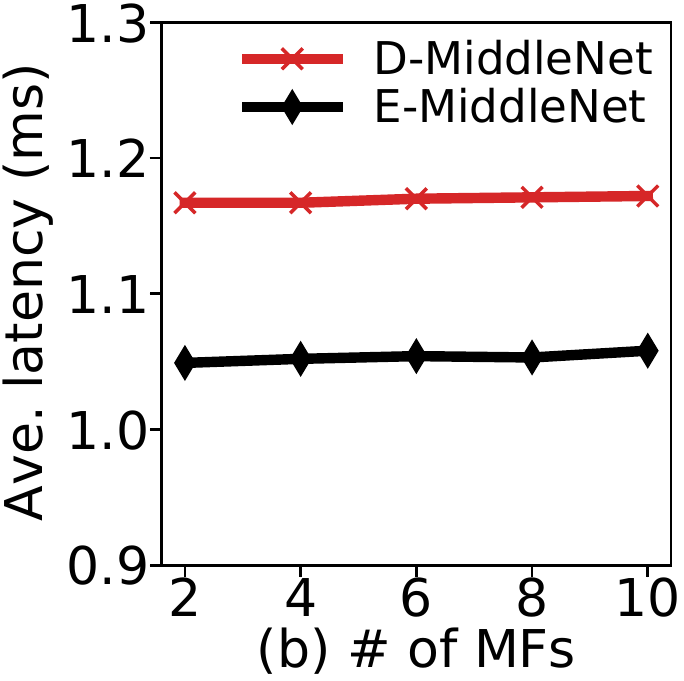}\label{fig:scale-latency}}\hfill
 \subfigure{\includegraphics[width=0.32\columnwidth]{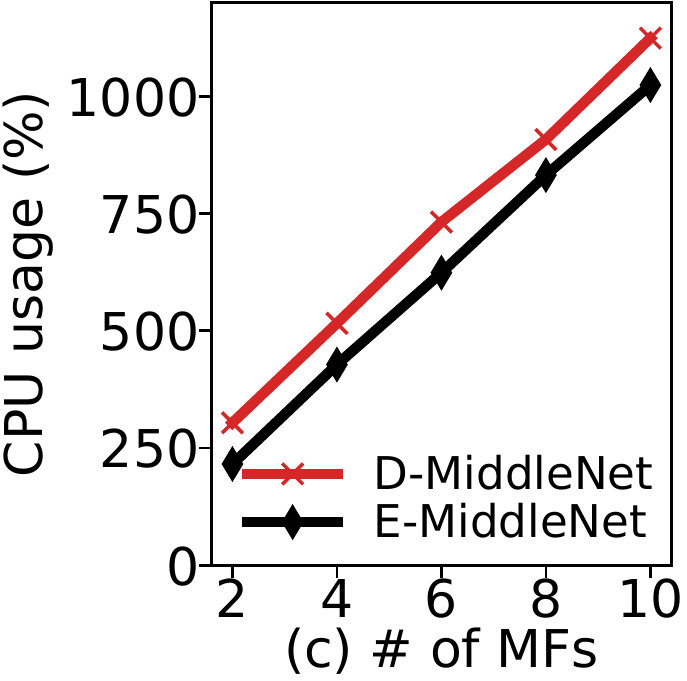}\label{fig:scale-cpu}}
\vspace{-3mm}
\caption{RPS (a), latency (b) and total CPU usage (c) comparison with increasing number of CPU-intensive MFs in the chain.}
\label{fig:scale-perf}
\end{figure}

\vspace{-1mm}\subsection{Performance Evaluation of L4/L7 middleboxes}\label{sec:l4l7-setup}\vspace{-1mm}
\noindent\textbf{Experiment Setup:} We now study the performance differences between the eBPF-based L4/L7 \name (Fig.~\ref{fig:middleboxes}, hereafter referred to as \eBPF) and the DPDK-based L4/L7 \name implementation (Fig.~\ref{fig:dpdk-middleboxes}, hereafter referred to as \dpdk).
As a third alternative, we use an NGINX proxy to study the impact of the loosely-coupled function chain (thus supporting a microservices paradigm) design in \name. The NGINX proxy acts as a non-virtualized proxy to perform functions
via internal function calls, which avoids introducing context switches or interrupts to achieve good data plane performance with a static, monolithic function implementation.
We also use the data plane model in
Fig.~\ref{fig:ovs-datapath} (d) (hereafter referred to as \textit{K-vSwitch}), as an additional alternative to compare with. 
We choose the Linux bridge as the implementation of the kernel-based vSwitch in Fig.~\ref{fig:ovs-datapath} (d). While the in-kernel OVS bridge could be another option, the Linux bridge offers all the functionality of a vSwitch for our evaluation purposes and is natively supported in Linux. In addition, the performance difference between Linux bridge and the in-kernel OVS bridge is not considered to be significant~\cite{netperf-nfv, netperf-microsvc}. It has also been noted that the in-kernel OVS bridge has difficulties being maintained as a separate project in addition to Linux kernel~\cite{ovs-afxdp-sigcomm}. 
We reuse most of the testbed setup described in \S\ref{sec:l2l3-setup}. 

We consider a typical HTTP workload (Apache Benchmark~\cite{apache-benchmark})
and examine application-level metrics, including request rate, response latency, and CPU usage, where the middlebox acts as a reverse proxy for web servers. 
The 1st node is configured to generate HTTP workloads.
The 2nd node is configured with the \name system. On the 3rd node, we configure two NGINX~\cite{nginx} instances as web servers. 
We enable adaptive batching for \eBPF to minimize the overhead incurred by frequent \skmsg interrupts within the chain 
at high concurrency.
We use a chain with two MFs. The first is a reverse proxy function that performs round-robin load balancing between the two web server backends on the 3rd node. 
The second function is a URL rewrite function that helps perform redirection for static websites.

We also compare the scalability of \dpdk and \eBPF, when the number of MFs in a linear chain increases. To evaluate the impact of CPU-intensive tasks on the network performance of MF chains, we let MFs perform prime number generation (based on the sieve-of-Atkin algorithm~\cite{sieve-of-atkin})
when a request is received. Each MF is assigned one dedicated CPU core to perform tasks, including RX/TX of requests and the prime number generation. We set the concurrency level (\ie the number of clients sending HTTP requests concurrently) of Apache Benchmark to 512 to generate sufficient load.

\noindent\textbf{Evaluation:} Fig.~\ref{fig:l4l7-perf} compares the RPS, response latency, and CPU usage of the different alternatives. \textit{K-vSwitch} has the lowest performance and highest CPU usage compared to the others. At a concurrency level of 512, the RPS of \textit{K-vSwitch} is only $\sim$42\% of the others, while its latency is $\sim$2.3$\times$ higher. The CPU usage of \textit{K-vSwitch} is even higher than \dpdk for concurrency levels greater than 16.  This demonstrates the heavyweight nature of the service function chain as discussed in \S\ref{sec:auditing} and demonstrates the benefit of having a zero-copy function chain (Appendix~\ref{sec:shm-auditing}) of the \name alternatives.

The use of \skmsg in \eBPF leads to slightly worse latency and throughput than \dpdk.
When the concurrency is between 1 and 32, there is a throughput difference between \dpdk and \eBPF,
ranging from 1.09$\times$ to 1.3$\times$.
At the lowest concurrency level of 1,
\eBPF consumes 37\% of the CPU, which is a 10$\times$ reduction compared to \dpdk (404\%, \ie 4 CPU cores). Since \dpdk uses polling to deliver packet descriptors, it continuously consumes CPU resources even when the traffic load is low, resulting in wasted CPU resources. Although \dpdk achieves 1.3$\times$ better RPS and latency compared to the \eBPF at a concurrency of 1, \eBPF's resource efficiency more than makes up for its lower throughput (which is likely not the goal when using a concurrency of 1, in any case) compared to \dpdk's constant usage of CPU. Thus, it is more  
desirable to use the lightweight \eBPF approach for these light loads.  

When the concurrency level increases and the load is higher, the adaptive batching of the \eBPF approach amortizes the interrupt and context switch overheads. 
The performance gap between \eBPF and the others reduces to be within 1.05$\times$ for concurrency levels higher than 64. 
With adaptive batching, \skmsg can pass a set of packet descriptors, incurring only one context switch and interrupt, saving substantial CPU cycles, reducing latency, and 
improving throughput. 

Compared to a monolithic NGINX as a middlebox, the \eBPF approach exhibits slightly worse throughput and latency performance (1.04$\times$ less RPS due to 1.04$\times$ higher response delay) because of the overhead of function chaining, \skmsg, and virtualization. 
NGINX's internal function calls have slightly lower overhead (25\% less on average) than 
\skmsg, which has additional context switches and interrupts. 
However, running a set of middleboxes as microservices improves flexibility and resiliency, allowing us to scale better, according to traffic load, especially with heterogeneous functions.
Moreover, it allows functions to be shared between different middlebox chains to improve resource utilization. With orchestration engines, \eg Kubernetes, intelligent scaling and placement policies can be applied with \name to improve resource efficiency further while still maintaining performance very close to a monolithic middlebox design.

Fig.~\ref{fig:scale-perf} evaluates the scalability of \dpdk and \eBPF with CPU-intensive MFs. Both \dpdk and \eBPF show good scalability as the number of MFs increases. Surprisingly, \eBPF performs even better than \dpdk with CPU-intensive MFs, with a 10\% improvement in RPS and a 10\%  reduction in latency. 
This is because with the prime number generation being CPU-intensive, it can quickly saturate the assigned CPU core and contend for CPU with the polling-based RX tasks of \dpdk's MF.
But for \eBPF, the RX of requests is triggered by interrupts, which is strictly load-proportional and avoids CPU contention. Since the prime number generation is performed within \eBPF's MFs, it is able to fully utilize the assigned CPU core, improving its performance.
To improve \dpdk's performance, more CPU resources need to be assigned to the MFs, meaning that we are using resources inefficiently. 
In addition, for the combined CPU usage of the message broker and MFs, \dpdk always needs one more CPU core than \eBPF (Fig.~\ref{fig:scale-cpu}). The extra CPU usage of \dpdk is due to the RX polling in the message broker to receive requests from the MF. Since prime number generation is time-consuming, it results in a lower request rate. This means that the CPU devoted to handling RX of requests is used inefficiently. This reiterates the fact that \dpdk uses resources inefficiently for this case, when dealing with CPU-intensive functions.

Throughout these experiments, \eBPF has significant resource savings at different concurrency levels compared to \dpdk, while having comparable throughput.
Further, \eBPF can even achieve better performance than \dpdk when it executes CPU-intensive functions even when it uses resources more frugally. 
It also achieves close to the same performance as a highly optimized, monolithic application like NGINX.  The resource efficiency benefits of the event-driven capability of eBPF, in conjunction with \skmsg to support shared memory processing, is a highly 
desirable way of building L4/L7 middlebox functionality in software.

\vspace{-1mm}\section{A Unified Design based on SR-IOV}\label{sec:unified}\vspace{-1mm}

Based on the understanding from studying the alternative approaches and their performance characteristics, we now develop the overall architecture of \name that supports the co-existence of network resident NFV and middlebox capabilities in a unified framework running on a single system. 

\begin{figure}[b]
\vspace{-4mm}
\centering
    \includegraphics[width=\columnwidth]{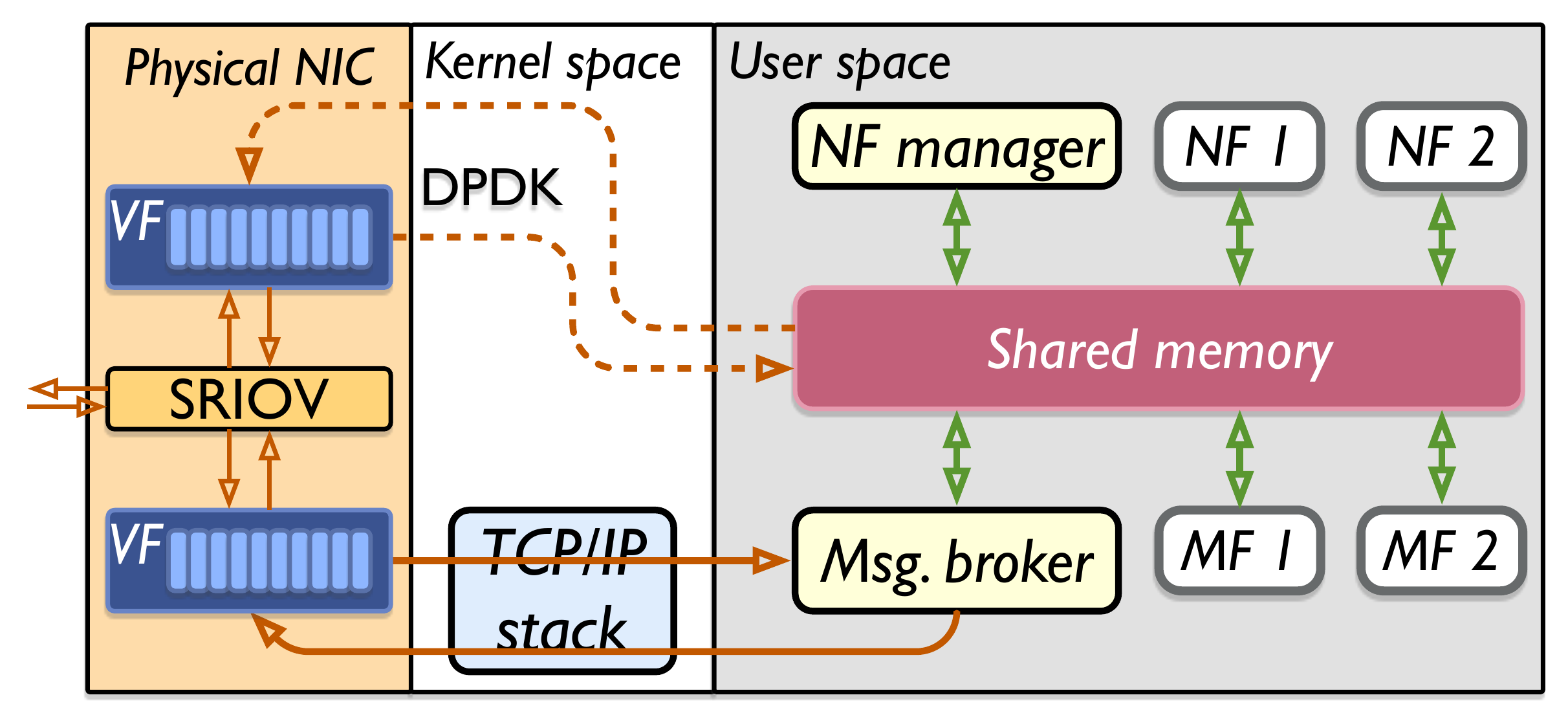}\vspace{-3mm}
\caption{The overall architecture of \name: A Combination of DPDK and eBPF via SR-IOV.}
\label{fig:framework}
\end{figure}

SR-IOV~\cite{Dong2010SRIOV} allows multiple Virtual Functions (VFs) on a shared NIC, as depicted in Fig.~\ref{fig:framework}. A VF acts as a distinct logical interface on the PCIe that offers direct access to the physical NIC resources that are shared across multiple VFs. It still 
achieves close to the single physical NIC's performance.
By dividing the hardware resources available on the physical NIC into multiple VFs, we can dedicate a VF for each L2/L3 \name and L4/L7 \name without having anyone take up the entire physical NIC.
The aggregate {\bf NIC} performance will still be at the line rate. 
\name uses the Flow Bifurcation mechanism~\cite{Flow-Bifurcation} for splitting traffic within the physical NIC 
in a flow or state-dependent manner. Since each VF is associated with different IP and MAC addresses, \name dynamically selects the packet processing layer (based on the VF it is attached to) from L2 to L7, providing a rich set of network-resident capabilities.

\vspace{-1mm}\subsection{Flow and State-dependent packet processing using SR-IOV}\vspace{-1mm}

\name attaches 
flow rules to the packet classifier in the physical NIC to support flow (and possibly state) dependent packet processing. Once a packet is received, the packet classifier parses and processes it based on its IP 5-tuple (\ie source/destination IPs, source/destination ports, protocol), which helps differentiate between packet flows. 
\newline\noindent\textbf{(1)} For a packet that needs to be handled by L2/L3 NFs, the classifier hands it to the VF bound to DPDK. The VF DMA's the raw packet to the shared memory in userspace. On the other side, the NF manager obtains the packet descriptor via the PMD and processes the packet in shared memory.
\newline\noindent\textbf{(2)} For a packet that needs to be handled by L4/L7 middlebox functions (MFs), the packet classifier 
hands the packet to the kernel TCP/IP stack through the corresponding VF. Since L4/L7 MFs require transport layer processing, 
\name utilizes the 
full-featured kernel protocol stack.

Because SR-IOV allows multiplexing of physical NIC resources, the split between the DPDK path and Linux kernel protocol stack path can be easily handled.
L2/L3 NFs and L4/L7 MFs can co-exist on the same node in \name.

Using SR-IOV in a simple design, however, would result in these two frameworks co-existing as two distinct and separate functions providing services for distinct flows.
There are two options for bridging the L2/L3 \name and L4/L7 \name: (1) A hardware-based approach that utilizes the NIC switch feature offered by SR-IOV~\cite{nic-switches} to connect different VFs within the NIC;\footnote{A SR-IOV enabled NIC must include the internal hardware bridge to support forwarding and packet classification between VFs on the same NIC.} 
(2) A software-based approach that uses \textit{virtio-user/vhost-net} \& \textit{TUN/TAP} device interfaces to connect L2/L3 \name to the kernel stack (see Fig.~\ref{fig:ovs-datapath} (b)), which is then connected to L4/L7 \name.\footnote{DPDK's Kernel NIC Interface (KNI~\cite{dpdk-kni}) is another software-based approach that provides equivalent functionality as \textit{virtio-user/vhost-net} \& \textit{TUN/TAP}. However, KNI lacks several important features compared to \textit{virtio-user/vhost-net} \& \textit{TUN/TAP}, such as multi-queue support, checksum offloading, \etc This makes the performance of KNI not as comparable as \textit{virtio-user/vhost-net} \& \textit{TUN/TAP}~\cite{virtio-user}.}

\begin{table}[t]
\centering
\caption{Overhead auditing of unified designs}\vspace{-3mm}
\label{tab:unified}
\resizebox{\columnwidth}{!}{%
\begin{tabular}{|c||c|c|}
\hline
 & NIC switch in SR-IOV & \begin{tabular}[c]{@{}c@{}}\textit{virtio-user/vhost-net} \\ \& \textit{TUN/TAP}\end{tabular} \\ \hline
\# of interrupts     & 2 & 2 \\ \hline
\# of copies         & 1 & 2 \\ \hline
\# of context switch & 1 & 2 \\ \hline
\end{tabular}%
}
\vspace{-4mm}
\end{table}

Table~\ref{tab:unified} compares the overhead generated by different alternatives. We only audit the datapath overhead between the NF manager in L2/L3 and the message broker in L4/L7, as they are the entry point of L2/L3 and L4/L7 \name. The hardware-based approach 
seamlessly works with the kernel-bypass in L2/L3 \name and moves the packet from the L2/L3 \name to the NIC via DMA. The NIC switch forwards the packet to the VF attached to the kernel stack without incurring any CPU overhead. All the overhead in the hardware-based approach is caused by passing the packet from the kernel stack to the message broker, however, is still less than software-based approach. The software-based approach inevitably introduces extra overhead and may compromise the performance gain achieved by L2/L3 kernel bypass.
Based on the overhead auditing, we decide to use the NIC switch to have packets pass through the kernel protocol stack in or out of the L4/L7 layer to the L2/L3 NF, for both L2/L3 NFs and L4/L7 MFs to operate on the same flow.

\vspace{-1mm}\subsection{Performance evaluation of unified design}\vspace{-1mm}
We investigate the performance of a unified L2/L3 NFV and L4/L7 middlebox and examine the interaction between the two, using 
SR-IOV to split the traffic. 
To mitigate interference between the load generators for L2/L3 (Pktgen~\cite{Pktgen}) and L4/L7  (Apache Benchmark~\cite{apache-benchmark}), we deploy Pktgen on the 1st node and Apache Benchmark on the 3rd node. We configure two NGINX servers on the 3rd node as the L4/L7 traffic sink.
We configure two VFs on the 2nd node with SR-IOV and bind L2/L3 \name (DPDK) and L4/L7 \name (eBPF) to separate VFs.  We use the same NFs (L3 routing and L2 forwarding) and MFs (reverse proxy and URL rewrite) on the 2nd node as described in \S\ref{sec:l2l3-setup} and \S\ref{sec:l4l7-setup}. We modify the NFs and MFs to perform hairpin routing: L2/L3 NFs return traffic to the 1st node, and L4/L7 MFs return traffic to the 3rd node. Thus, we eliminate the interference that occurs between the two traffic generators. 
For L2/L3 traffic, we keep the sending rate at the MLFR. For L4/L7 traffic, we use a concurrency of 256 with the Apache Benchmark.

\begin{figure}[t]
\centering
    \subfigure{\includegraphics[width=0.49\columnwidth]{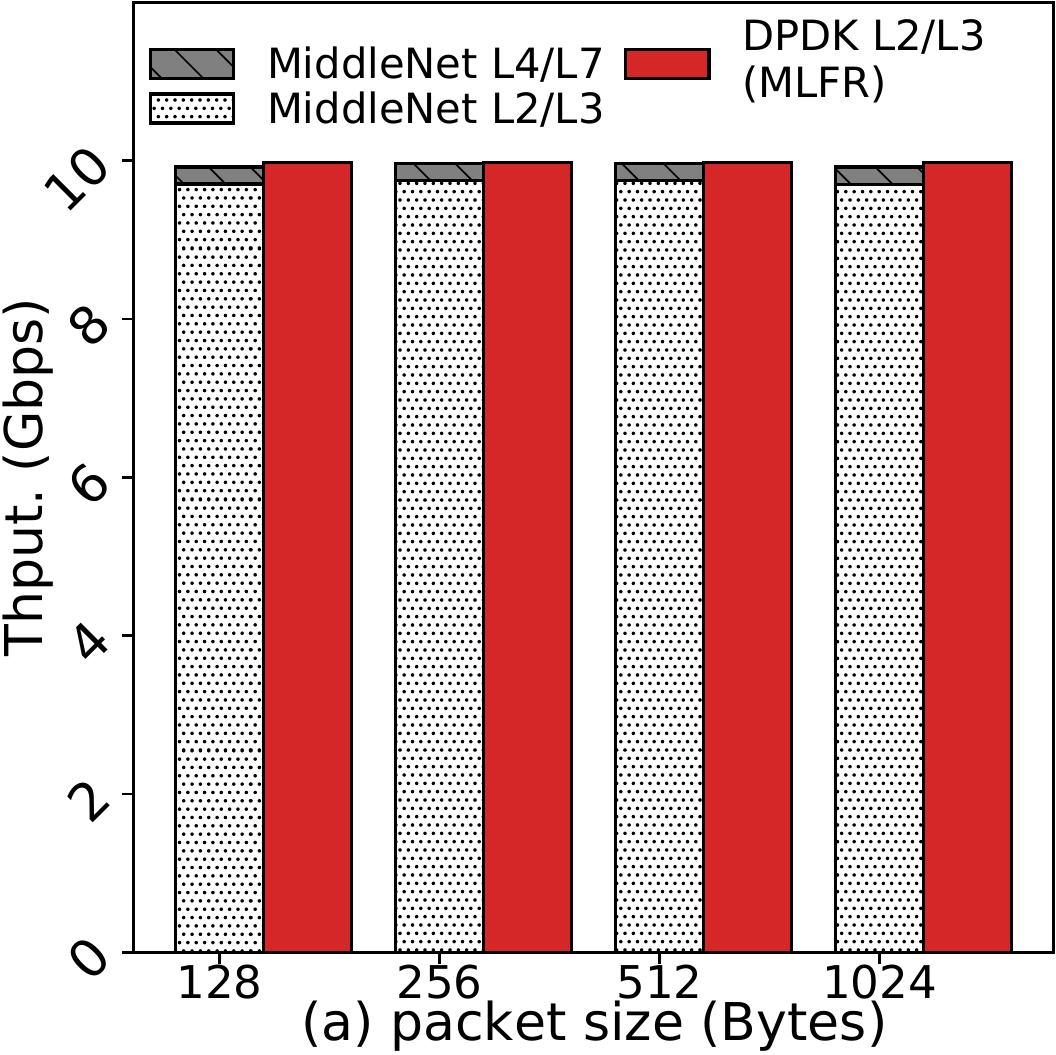}\label{fig:sr-iov-thput}}\hfill
    \subfigure{\includegraphics[width=0.49\columnwidth]{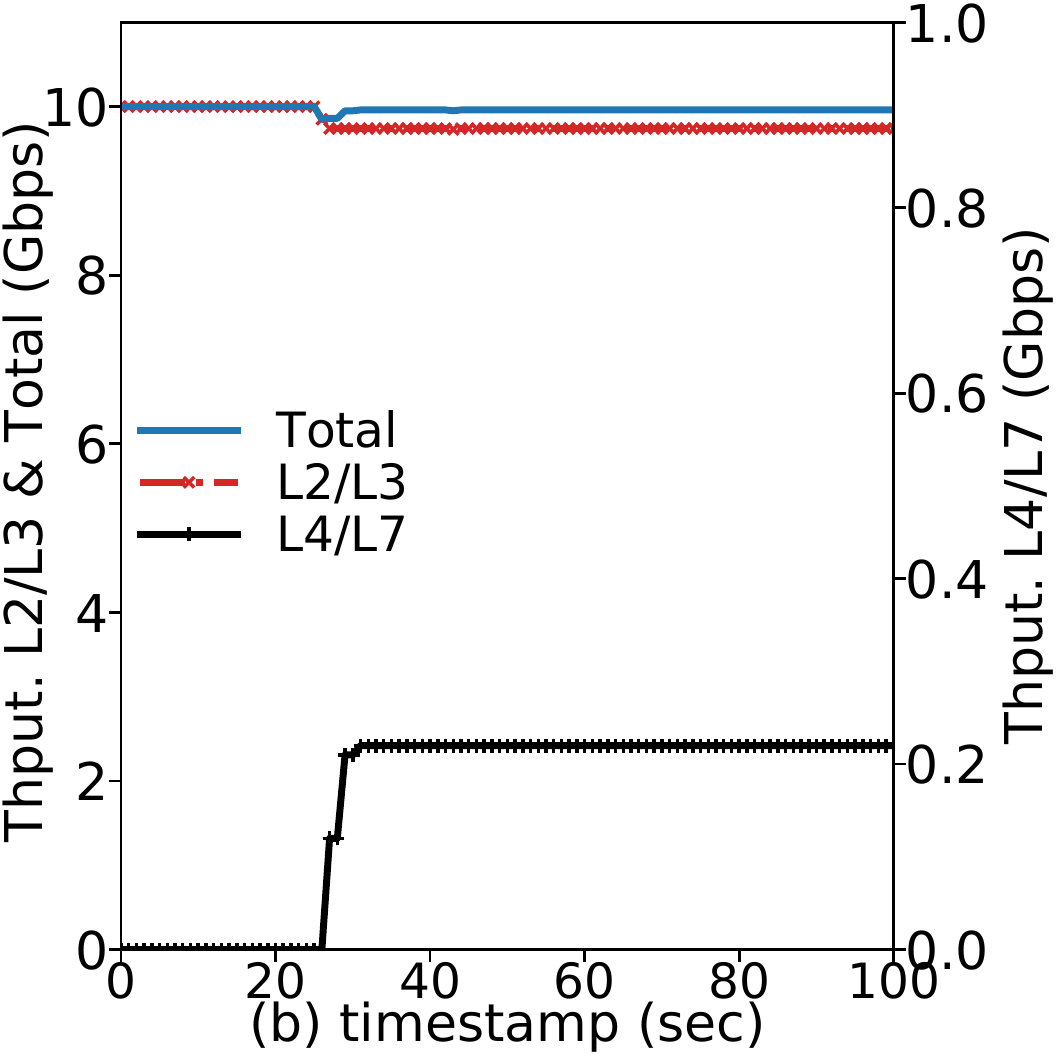}\label{fig:sr-iov-time}}
\vspace{-3mm}
\caption{(a) Aggregate throughput for various packet sizes. For L2/L3 NFV, we use Maximum loss free rate (MLFR) (b) Time series of throughput for L2/L3 NFV and total (left Y-axis) and L4/L7 middlebox (right Y-axis).}
\label{fig:sr-iov}\vspace{-4mm}
\end{figure}

We study whether there is interference by checking the aggregate throughput as well as the throughput for the L2/L3 traffic processed by NFV and the L4/L7 processed by the middlebox,
as shown in Fig.~\ref{fig:sr-iov-thput}. The aggregate throughput of L2/L3 NFs and L4/L7 MFs remains close to  10Gbps, with negligible performance loss across various packet sizes. 
We also study the impact of adding L4/L7 flows when L2/L3 traffic (128Bytes packets) goes through \name at line rate (10 Gbps link). As shown in Fig.~\ref{fig:sr-iov-time}, at the 25th second, the Apache Benchmark starts to generate L4/L7 traffic (0.22Gbps), and the throughput of L2/L3 NFs correspondingly drops  to 9.78Gbps.
Thus, our unified design in \name for the co-existence of DPDK-based L2/L3 NFs and eBPF-based MFs provides both flexibility 
and performance.

\vspace{-1mm}\section{Isolation and Security Domains in \name}\label{sec:security-domain}\vspace{-1mm}

The use of shared memory raises concerns as it may weaken the isolation/security boundary between the functions that share the same memory region. Our trust model assumes that only functions in \name trust each other. 
Functions in \name (NFs or MFs), which run as DPDK secondary processes, share the same private memory pool by using the same ``shared data file prefix'' (specified by the shared memory manager (\S\ref{sec:l2l3-overview})) during their startup. We `admission control' functions by validating the creation of a \name function that is authenticated and uses the correct file prefix. We additionally apply inter-function packet descriptor filtering to prevent unauthorized access to the data in shared memory, through the virtual address in the packet descriptor. In accordance with the way packet descriptors are passed, these are different for L2/L3 (with DPDK's RTE ring) \name versus  L4/L7 (with eBPF's \skmsg) \name.

\noindent\textbf{Descriptor filtering for L2/L3 NFs:}
We leverage the NF manager in L2/L3 \name to perform packet descriptor filtering. Once the NF manager polls a new packet descriptor from an NF's TX ring, it queries its internal filtering map and checks whether the packet descriptor is authorized to be sent to the target NF based on matched rules. Unauthorized packet descriptors are dropped by the NF manager.

\noindent\textbf{Descriptor filtering in L4/L7:}
Since the L4/L7 \name uses \skmsg to pass packet descriptors between functions (\S\ref{sec:skmsg-l4l7}), it is natural to exploit eBPF's extensibility to filter packet descriptors. We add an additional eBPF map to the \skmsg program to store filtering rules. Each time a packet descriptor arrives, the \skmsg program parses the destination of the packet descriptor and uses it as the key to lookup the filtering rule. The packet descriptor is passed to the destination if allowed; otherwise, the descriptor is recognized as unauthorized and discarded.

\vspace{-1mm}\section{Related Work}\vspace{-1mm}
NFV platforms use different implementation approaches and primarily operate at L2/L3. OpenNetVM~\cite{opennetvm}, based on DPDK, uses the microservice paradigm with a flexible composition of functions and uses shared memory to achieve full line-rate performance. However, OpenNetVM lacks full-fledged protocol stack support, focusing on supporting L2/L3 NFs. Compared to OpenNetVM, \name supports processing across the entire protocol stack, including application support. Other NFV platforms take different approaches. Both ClickOS~\cite{Joao2014ClickOS} and  NetMap~\cite{Rizzo2012Netmap} use traditional kernel style processing and mapping of kernel-user space memory, using interrupts for notifications. The interrupt-based notification schemes of ClickOS and NetMap can be vulnerable to poor overload behavior because of receive-livelocks~\cite{mogul1997eliminating}. In contrast, the L2/L3 processing in \name uses polling, thus avoiding receive-livelocks. E2~\cite{Palkar2015E2} integrates all the NFs as one monolith to help improve performance but gives up some flexibility to build complex NF chains through the composition of independently developed NFs. NFV designs have increasingly adopted the microservice paradigm for flexible composition of functions while still striving to achieve full line-rate performance. Supporting this, \name's disaggregated design offers the flexibility to build complex L2/L3 NF chains. 

Network-resident middleboxes' functionality depends on having full kernel protocol processing, typically terminating a transport layer connection and requiring a full-fledged protocol stack. Efforts have been made to pursue a high-performance middlebox framework with protocol processing support~\cite{Muhammad2017mOS, Liu2018Microboxes, Kenichi2016StackMap}. However, each of these proposals has its difficulties. mOS~\cite{Muhammad2017mOS} focuses on developing a monolithic middlebox, lacking the flexibility of a disaggregated design like \name. Microboxes~\cite{Liu2018Microboxes} leverages DPDK and OpenNetVM's shared memory design to improve packet processing performance and achieve flexible middlebox chaining. However, it does not provide a full-fledged protocol stack (it only supports TCP). The CPU consumption of DPDK-based designs is a further deterrent in the L4/L7 use case, significantly when the chain's complexity increases. Establishing communication channels for a chain of middleboxes using the kernel network stack incurs considerable overhead. Every transfer between distinct middleboxes typically involves full protocol stack traversals, which adds considerable overhead. It typically involves \textit{two} data copies, context switches, protocol stack processing, multiple interrupts,  and \textit{one} serialization and deserialization operation. \name is designed to reduce these overheads by leveraging shared memory processing, in the meanwhile, adopting eBPF-based event-driven processing to minimize CPU consumption. StackMap~\cite{Kenichi2016StackMap}  also leverages the  feature-rich kernel protocol stack to perform protocol processing while bypassing the kernel to improve packet I/O performance. However, it is more focused on end-system support than middlebox function chaining. StackMap's capability may be complementary to the design of \name.

There has not been a significant effort to design a unified environment where L2/L3 NFV and L4/L7 middlebox environments co-exist.  \name is designed to address this issue.

\noindent\textbf{eBPF-based NFV/Middlebox:}
\cite{Miano2021Polycube, abranches2022getting, van2020accelerating} explore the use of eBPF to implement NFV/Middlebox functions. These eBPF-based functions reside in the kernel, running as a set of eBPF programs attached at various eBPF hooks, \eg eXpress Data Path (XDP), and Traffic Control (TC). This avoids expensive context switches, as packet processing always remains within the kernel. In addition, since the packet payload is retained in the kernel buffers. Only the packet metadata,\footnote{The packet metadata is represented as a ``xdp\_md'' data structure when using the XDP hook, and is in the form of a ``sk\_buff'' data structure when using TC hook.} which contains packet descriptor, is passed between different eBPF-based functions, thus achieving zero-copy packet delivery in the kernel. Compared to \name, \cite{Miano2021Polycube, abranches2022getting, van2020accelerating} focus on the affinity in the kernel. In contrast, L2/L3 \name relies on DPDK, which uses SR-IOV to achieve a unified design. \cite{Miano2021Polycube, abranches2022getting, van2020accelerating} can seamlessly work with the kernel protocol stack for protocol processing.
However, the eBPF-based functions in \cite{Miano2021Polycube, abranches2022getting, van2020accelerating} are triggered using kernel interrupts, thus potentially suffering from poor  overload behavior~\cite{mogul1997eliminating}. Thus, their approach can perform poorly compared to L2/L3 \name, which leverages DPDK to achieve line-rate performance. 
Additionally, the eBPF-based functions can only be used to support L2/L3/L4 use cases within the kernel. Since L7 middleboxes not only require protocol processing, but have application code that typically run in userspace, approaches as in~\cite{Miano2021Polycube, abranches2022getting, van2020accelerating} result in expensive packet transfers between the kernel performing packet processing and the L7 userspace application.
The shared memory design in L4/L7 \name avoids this overhead,
thus achieving better data plane performance for a unified L4/L7 environment.

\vspace{-1mm}\section{Conclusion}\vspace{-1mm}

We presented \name, a unified environment supporting L2/L3 NFV functionality and L4/L7 middleboxes. In \name, we chose the high-performance packet processing of DPDK for L2/L3 NFs and the resource efficiency of eBPF for L4/L7 middlebox functions. \name leverages shared memory processing for both use cases to support high-performance function chains. 
Experimental results demonstrated the performance benefits of using DPDK for L2/L3 NFV. \name can achieve full line rate for almost all packet sizes given adequate CPU resources provided to \name's NF manager. Its throughput outperforms an eBPF-based design that depends on interrupts by 4$\times$ 
for small packets and has a 2$\times$ reduction in latency. For the L4/L7 use case, the performance of our eBPF-based design in \name is close to the DPDK-based approach, getting
to within 1.05$\times$ at higher loads (large concurrency levels).
In addition, the eBPF-based approach has significant resource savings, with an average of 3.2$\times$ reduction in CPU usage compared to a DPDK-based L4/L7 design.
Using SR-IOV on the NIC, \name creates a unified environment with negligible impact on performance, 
running the DPDK-based L2/L3 NF chains and eBPF-based L4/L7 middlebox chains on the same node. This can bring substantial deployment flexibility.

\section*{Acknowledgments}\vspace{-1mm}
We thank US National Science Foundation for their generous support through grants CRI-1823270 and CSR-1763929.

\appendices

\section{Details of DPDK's shared memory support}\label{appendix:dpdk-shm-support}

After the DPDK primary process (\ie shared memory manager)  initializes the memory pools, it writes the memory pool information (\eg base virtual address, the allocated huge pages) into a configuration file through DPDK's EAL (Environment Abstraction Layer~\cite{dpdk-eal}). The DPDK secondary processes (\ie functions, L2/L3 NF manager, L4/L7 message broker) read the configuration file during startup and use DPDK's EAL to map the same memory regions allocated by the DPDK primary process. This ensures all the DPDK secondary processes share the same memory pools, thereby facilitating shared memory communication between functions.

When VMs are used, they rely on the emulated PCI to access physical memory in the host. This requires multiple address translations (\ie Guest Virtual Address to Guest Physical Address and then to Host Virtual Address). This adds a burden while sharing memory across different VMs, since they have different virtual address mappings to the host. It requires the hypervisor (as it knows the virtual address mappings of different VMs) to remap the base virtual address in the packet descriptor, which adds additional processing latency. In contrast, a container shares the same virtual memory address, which means that its virtual address can be interpreted by other containers without an additional translation. This facilitates memory sharing between different functions implemented in containers and makes it straightforward to build shared memory for function chains using existing tools such as DPDK's multi-process support.

\section{Overhead auditing of function chains using shared memory}\label{sec:shm-auditing}\vspace{-1mm}
To \textit{quantitatively} understand the benefit of shared memory communication and the difference between alternatives, we now perform an auditing of the overheads for the function chain in Fig.~\ref{fig:shm-pipeline}.

\begin{table}[b]
\vspace{-5mm}
\centering
\caption{Overhead auditing of L2/L3 NF chain using shared memory communication}
\label{tab:l2l3-shm}\vspace{-3mm}
\resizebox{\columnwidth}{!}{%
\begin{tabular}{|cc|cc|cccc|c|}
\hline
\multicolumn{2}{|c|}{\multirow{2}{*}{\textbf{Data pipeline No.}}} &
  \multicolumn{2}{c|}{NIC-shared memory} &
  \multicolumn{4}{c|}{Within the chain} &
  \multirow{2}{*}{\textbf{total}} \\ \cline{3-8}
\multicolumn{2}{|c|}{} &
  \multicolumn{1}{c|}{\quad\ding{172}\quad\quad} &
  \ding{177} &
  \multicolumn{1}{c|}{\ding{173}} &
  \multicolumn{1}{c|}{\ding{174}} &
  \multicolumn{1}{c|}{\ding{175}} &
  \ding{176} &
   \\ \hline\hline
\multicolumn{1}{|c|}{\multirow{2}{*}{\textbf{\# of copies}}} &
  ($\alpha$) polling &
  \multicolumn{1}{c|}{0} &
  0 &
  \multicolumn{1}{c|}{0} &
  \multicolumn{1}{c|}{0} &
  \multicolumn{1}{c|}{0} &
  0 &
  0 \\ \cline{2-9} 
\multicolumn{1}{|c|}{} &
  ($\beta$) event-driven &
  \multicolumn{1}{c|}{0} &
  0 &
  \multicolumn{1}{c|}{0} &
  \multicolumn{1}{c|}{0} &
  \multicolumn{1}{c|}{0} &
  0 &
  0 \\ \hline\hline
\multicolumn{1}{|c|}{\multirow{2}{*}{\textbf{\# of interrupts}}} &
  ($\alpha$) polling &
  \multicolumn{1}{c|}{0} &
  0 &
  \multicolumn{1}{c|}{0} &
  \multicolumn{1}{c|}{0} &
  \multicolumn{1}{c|}{0} &
  0 &
  0 \\ \cline{2-9} 
\multicolumn{1}{|c|}{} &
  ($\beta$) event-driven &
  \multicolumn{1}{c|}{2} &
  1 &
  \multicolumn{1}{c|}{1} &
  \multicolumn{1}{c|}{1} &
  \multicolumn{1}{c|}{1} &
  1 &
  7 \\ \hline\hline
\multicolumn{1}{|c|}{\multirow{2}{*}{\textbf{\# of context switch}}} &
  ($\alpha$) polling &
  \multicolumn{1}{c|}{0} &
  0 &
  \multicolumn{1}{c|}{0} &
  \multicolumn{1}{c|}{0} &
  \multicolumn{1}{c|}{0} &
  0 &
  0 \\ \cline{2-9} 
\multicolumn{1}{|c|}{} &
  ($\beta$) event-driven &
  \multicolumn{1}{c|}{1} &
  1 &
  \multicolumn{1}{c|}{1} &
  \multicolumn{1}{c|}{1} &
  \multicolumn{1}{c|}{1} &
  1 &
  6 \\ \hline
\end{tabular}%
}
\vspace{.5ex}

{\raggedright
($\alpha$) \textit{polling}-based kernel-bypass (using DPDK's PMD) + \textit{polling}-based zero-copy I/O for function chaining (using DPDK's RTE RING); \\
($\beta$) \textit{event-driven} kernel-bypass (using eBPF's AF\_XDP) + \textit{event-driven} zero-copy I/O for function chaining (using eBPF's \skmsg). \par}

\end{table}

\noindent\textit{(1) L2/L3 NF use case:}
For the L2/L3 NF use case, we study two alternatives: first is ($\alpha$) NIC-shared memory packet exchange with \textit{polling}-based kernel-bypass (using DPDK's PMD) + \textit{polling}-based zero-copy I/O for function chaining (using DPDK's RTE RING); second is ($\beta$) NIC-shared memory packet exchange with \textit{event-driven} kernel-bypass (using eBPF's AF\_XDP) + \textit{event-driven} zero-copy I/O for function chaining (using eBPF's \skmsg).
We skip the \textit{kernel-based} NIC-shared memory packet exchange in this auditing, as it is apparently unsuitable for L2/L3 NFs.

Table~\ref{tab:l2l3-shm} shows the overhead auditing of L2/L3 NF scenario for both (($\alpha$) and ($\beta$)). 
Compared to the optimal L2/L3 data plane model (f) discussed in \S\ref{sec:auditing}, the polling-based shared memory communication approach ($\alpha$) avoids any data copy, interrupt, and context switch, throughout the entire data pipeline (from \ding{172} to \ding{177} of Fig.~\ref{fig:shm-pipeline}). The event-driven alternative ($\beta$) eliminates all the data copies as well. However, the use of AF\_XDP and \skmsg introduces additional interrupts and context switches. In particular, every packet transfer within the chain incurs one interrupt and context switch, which is a non-negligible overhead, especially if the chain grows in scale.

\begin{table}[b]
\vspace{-5mm}
\centering
\caption{Overhead auditing of L4/L7 middlebox chain using shared memory communication}
\label{tab:l4l7-shm}\vspace{-3mm}
\resizebox{\columnwidth}{!}{%
\begin{tabular}{|cc|cc|cccc|c|}
\hline
\multicolumn{2}{|c|}{\multirow{2}{*}{\textbf{Data pipeline No.}}} &
  \multicolumn{2}{c|}{NIC-shared memory} &
  \multicolumn{4}{c|}{Within the chain} &
  \multirow{2}{*}{\textbf{total}} \\ \cline{3-8}
\multicolumn{2}{|c|}{} &
  \multicolumn{1}{c|}{\quad\ding{172}\quad\quad} &
  \ding{177} &
  \multicolumn{1}{c|}{\ding{173}} &
  \multicolumn{1}{c|}{\ding{174}} &
  \multicolumn{1}{c|}{\ding{175}} &
  \ding{176} &
   \\ \hline\hline
\multicolumn{1}{|c|}{\multirow{2}{*}{\textbf{\# of copies}}} &
  ($\gamma$) polling &
  \multicolumn{1}{c|}{2} &
  2 &
  \multicolumn{1}{c|}{0} &
  \multicolumn{1}{c|}{0} &
  \multicolumn{1}{c|}{0} &
  0 &
  4 \\ \cline{2-9} 
\multicolumn{1}{|c|}{} &
  ($\delta$) event-driven &
  \multicolumn{1}{c|}{2} &
  2 &
  \multicolumn{1}{c|}{0} &
  \multicolumn{1}{c|}{0} &
  \multicolumn{1}{c|}{0} &
  0 &
  4 \\ \hline\hline
\multicolumn{1}{|c|}{\multirow{2}{*}{\textbf{\# of interrupts}}} &
  ($\gamma$) polling &
  \multicolumn{1}{c|}{2} &
  1 &
  \multicolumn{1}{c|}{0} &
  \multicolumn{1}{c|}{0} &
  \multicolumn{1}{c|}{0} &
  0 &
  3 \\ \cline{2-9} 
\multicolumn{1}{|c|}{} &
  ($\delta$) event-driven &
  \multicolumn{1}{c|}{2} &
  1 &
  \multicolumn{1}{c|}{1} &
  \multicolumn{1}{c|}{1} &
  \multicolumn{1}{c|}{1} &
  1 &
  7 \\ \hline\hline
\multicolumn{1}{|c|}{\multirow{2}{*}{\textbf{\# of context switch}}} &
  ($\gamma$) polling &
  \multicolumn{1}{c|}{1} &
  1 &
  \multicolumn{1}{c|}{0} &
  \multicolumn{1}{c|}{0} &
  \multicolumn{1}{c|}{0} &
  0 &
  2 \\ \cline{2-9} 
\multicolumn{1}{|c|}{} &
  ($\delta$) event-driven &
  \multicolumn{1}{c|}{1} &
  1 &
  \multicolumn{1}{c|}{1} &
  \multicolumn{1}{c|}{1} &
  \multicolumn{1}{c|}{1} &
  1 &
  6 \\ \hline\hline
\multicolumn{1}{|c|}{\multirow{2}{*}{\textbf{\begin{tabular}[c]{@{}c@{}}\# of protocol \\ processing tasks\end{tabular}}}} &
  ($\gamma$) polling &
  \multicolumn{1}{c|}{1} &
  1 &
  \multicolumn{1}{c|}{0} &
  \multicolumn{1}{c|}{0} &
  \multicolumn{1}{c|}{0} &
  0 &
  2 \\ \cline{2-9} 
\multicolumn{1}{|c|}{} &
  ($\delta$) event-driven &
  \multicolumn{1}{c|}{1} &
  1 &
  \multicolumn{1}{c|}{0} &
  \multicolumn{1}{c|}{0} &
  \multicolumn{1}{c|}{0} &
  0 &
  2 \\ \hline\hline
\multicolumn{1}{|c|}{\multirow{2}{*}{\textbf{\begin{tabular}[c]{@{}c@{}}\# of serialization \\ or deserialization (L7) \end{tabular}}}} &
  ($\gamma$) polling &
  \multicolumn{1}{c|}{1} &
  1 &
  \multicolumn{1}{c|}{0} &
  \multicolumn{1}{c|}{0} &
  \multicolumn{1}{c|}{0} &
  0 &
  2 \\ \cline{2-9} 
\multicolumn{1}{|c|}{} &
  ($\delta$) event-driven &
  \multicolumn{1}{c|}{1} &
  1 &
  \multicolumn{1}{c|}{0} &
  \multicolumn{1}{c|}{0} &
  \multicolumn{1}{c|}{0} &
  0 &
  2 \\ \hline
\end{tabular}%
}
\vspace{.5ex}

{\raggedright
($\gamma$) \textit{kernel-based} NIC-shared memory packet exchange + \textit{polling}-based zero-copy I/O for function chaining (using DPDK's RTE RING); \\
($\delta$) \textit{kernel-based} NIC-shared memory packet exchange + \textit{event-driven} zero-copy I/O for function chaining (using eBPF's \skmsg). \par}

\end{table}
\noindent\textit{(2) L4/L7 middlebox use case:}
For the L4/L7 middlebox use case, we study two alternatives: 
($\gamma$) \textit{kernel-based} NIC-shared memory packet exchange + \textit{polling}-based zero-copy I/O for function chaining (using DPDK's RTE RING);
($\delta$) \textit{kernel-based} NIC-shared memory packet exchange + \textit{event-driven} zero-copy I/O for function chaining (using eBPF's \skmsg). 
We skip the \textit{kernel-bypass} NIC-shared memory packet exchange in this auditing, as L4/L7 middleboxes depend on the kernel stack for protocol processing.

Table~\ref{tab:l4l7-shm} shows the overhead auditing of L4/L7 middlebox options (($\gamma$) and ($\delta$)). Compared to the optimal L4/L7 data plane model (d) in \S\ref{sec:auditing}, the polling-based ($\gamma$) and event-driven ($\delta$) shared memory communication approaches avoid any data copy within the function chain (\ding{173} to \ding{176} in Fig.~\ref{fig:shm-pipeline}), because of the zero-copy I/O.
However, moving a packet from the NIC to shared memory (\ding{172} in Table~\ref{tab:l4l7-shm}) incurs two data copies, and vice versa (\ding{177} in Table~\ref{tab:l4l7-shm}). One data copy comes from the packet exchange between the NIC and the message broker (Fig.~\ref{fig:shm-pipeline}), where the kernel stack needs to copy the packet from the kernel to the message broker in userspace, after protocol processing. The message broker then moves the packet into shared memory, which introduces the second copy. With the middlebox chain of two functions, using shared memory communication (($\gamma$) or ($\delta$)) shows no significant benefit compared to optimal L4/L7 data plane model (d) because of the data copy incurred when moving packets between the NIC and shared memory. They all introduce 4 data copies throughout the entire data pipeline (from \ding{172} to \ding{177} in Fig.~\ref{fig:shm-pipeline} and Fig.~\ref{fig:data-pipeline}). The shared memory communication for the L4/L7 middlebox scenario (($\gamma$), ($\delta$)) shows its advantages of saving on data copies (due to the zero-copy I/O) compared to the L4/L7 data plane model (d) only when the size of the chain grows. In comparison, the data copy overhead in (d) will increase as the chain increases.

Another essential asset of shared memory communication is that it completely eliminates protocol processing, serialization, and deserialization overheads within the chain. These tasks are performed before the packet is moved to shared memory by the message broker, and vice versa (\ding{172} and \ding{177} in Table~\ref{tab:l4l7-shm}). No matter the size of the chain, the total \# of protocol processing tasks or serialization/deserialization tasks incurred when using shared memory communication is always \textit{two}. On the other hand, these overheads in the data plane model (d) increase as the chain scales, indicating poor scalability.

The event-driven approach ($\delta$), which uses \skmsg to implement the zero-copy I/O, incurs  one interrupt and one context switch for each transmission within the function chain (\ding{173} to \ding{176} in Fig.~\ref{fig:shm-pipeline}).
This inevitably has a higher latency compared to using DPDK's RTE RING. With DPDK's RTE RING, different functions exchange packet descriptors entirely in userspace and avoid expensive context switches. 
For the I/O latency going from one function to the next,
eBPF’s \skmsg needs $\sim$20 microseconds to send each packet descriptor. On the other hand, DPDK’s RTE RING only needs $\sim$0.5 microseconds. This penalty with \skmsg's kernel interrupts and context switching overheads 
makes the low-latency DPDK’s RTE RING ideal for building high-performance function chains,  desirable for latency-sensitive workloads. 
However, DPDK’s RTE RING comes at the cost of constant polling and thus resource consumption.
From a resource efficiency standpoint, \skmsg’s event-driven nature makes it more efficient, because it does \textit{not} consume CPU cycles when there is no traffic. This is similar to AF\_XDP, as they both belong to the eBPF system of Linux. The latency of \skmsg is less of a concern if there are other dominant latencies masking it. This is often true for L4/L7 middleboxes, where application-level latency and kernel protocol processing latency dominate the total request delay.
It requires further optimization on the use of \skmsg, \eg having packet descriptors directly routed between functions without being mediated by the message broker (details in \S\ref{sec:skmsg-l4l7}), which can considerably reduce the amount of interrupt and context switch generated by \skmsg.

\bibliographystyle{IEEEtran}
\bibliography{myref}

\end{document}